\documentclass[a4paper,12pt]{article}
\pdfoutput=1

\usepackage{amsmath, amssymb, authblk, a4wide, bm, cancel, color, graphicx, subfig, enumerate, appendix, youngtab, cite, hyperref, bbold, verbatim, slashed}

\newcommand{\pdiff}[2]{\frac{\partial#1}{\partial#2}}

\newcommand{\nbrack}[1]{\left(#1\right)}
\newcommand{\sbrack}[1]{\left[#1\right]}

\newcommand{\norm}[1]{\left|#1\right|}
\newcommand{\pfrac}[2]{\left(\frac{#1}{#2}\right)}

\def\be{\begin{equation}} \def\ee{\end{equation}}
\def\ba{\begin{eqnarray}} \def\ea{\end{eqnarray}} 

\def\cL{{\cal L}}

\def\uno{\mbox{1 \kern-.59em {\rm l}}}

\numberwithin{equation}{section}
\numberwithin{figure}{section}
\numberwithin{table}{section}

\begin{document}
\title{\vspace{2cm}
\Large{\textbf{Collider constraints on tuning in composite Higgs models}}}
\author[1]{\small{\bf James Barnard}\thanks{\texttt{james.barnard@unimelb.edu.au}}}
\author[2]{\small{\bf Martin White}\thanks{\texttt{martin.white@adelaide.edu.au}}}
\affil[1]{ARC Centre of Excellence for Particle Physics at the Terascale, School of Physics, University of Melbourne, Victoria 3010, Australia}
\affil[2]{ARC Centre of Excellence for Particle Physics at the Terascale, School of Physics, University of Adelaide, South Australia 5005, Australia}
\date{}
\maketitle

\begin{abstract}
\baselineskip=15pt
\noindent 
Two potential sources of tuning exist in composite Higgs models: one comes from keeping the Higgs VEV below the compositeness scale and one comes from keeping the Higgs light after constraints on the top partner masses are applied. We construct a measure that determines whether these tunings are independent or not and combines them appropriately. We perform a comprehensive scan of the parameter space for three explicit models and report the minimum tuning values compatible with existing collider constraints. Tuning values are given as functions of resonance masses and deviations to the Higgs couplings so the effect of future constraints can easily be quantified. The current minimum tuning in the minimal model is 2.5-5\% and will be decreased to around 0.8-3.3\% if no top partners are observed over the lifetime of the LHC\@.
\end{abstract}

\newpage

\section{Introduction}

For a long time the concept of naturalness has provided strong motivation to expect new physics at the TeV scale. As yet, evidence for this new physics stubbornly remains unforthcoming and many extensions to the Standard Model are coming under increasing pressure. Whether they remain natural or not is becoming an ever more common question.

A particularly compelling way of extending the Standard Model is to replace the elementary Higgs with a composite one, specifically a pseudo Nambu-Goldstone boson of a new, spontaneously broken, global symmetry~\cite{Kaplan:1983fs,Kaplan:1983sm,Dugan:1984hq}. Not only is this an intriguing prospect by itself, given that we already know of several examples of compositeness in nature, it is also able to explain the fermion mass hierarchies~\cite{Kaplan:1991dc,Gherghetta:2000qt} and, of course, render the Standard Model natural, many composite Higgs sectors being insensitive to physics at scales above the compositeness scale. Regrettably the dearth of hints for new physics points towards a compositeness scale significantly higher than the electroweak scale, thus protection from physics at higher scales is not worth what it once was and most models reproducing the correct electroweak scale are expected to be tuned~\cite{Archer:2014qga}.\footnote{Tuned composite Higgs models may still be of interest for other phenomenological reasons~\cite{1409.7391}.} The main aims of this paper are to find out just how tuned composite Higgs models are now, and how this will change after future collider experiments.

Several factors contribute to the tuning in composite Higgs models. The most obvious is the tuning required to keep the Higgs VEV below the compositeness scale. Even then it has been known for a while now that, in the most commonly considered models, the Higgs mass is further correlated with the mass of the lightest top partner~\cite{Matsedonskyi:2012ym,Marzocca:2012zn,Pomarol:2012qf,Panico:2012uw,Pappadopulo:2013vca,Barnard:2013hka,1405.1498}, top partners being the coloured, composite fermions that allow the top quark to couple strongly to the Higgs. A consequence key to our study is that collider constraints on top partner masses cause a second tuning in the Higgs mass. To properly quantify the overall tuning in a given model we must therefore devise a tuning measure that determines whether any two tunings are independent, then combines them in a way that accounts for this.

The other major obstacle in answering our central questions comes from the number of parameters. In the simplest, viable model there are still nine parameters and, by design, we are searching for tuned regions of the parameter space. The machinations taking a parameter point to a Higgs VEV and mass are not easily invertible so we could be left with a costly scanning problem, especially if we want to be convinced that we have sufficient coverage of the parameter space for the tuning values we find to be representative. Our strategy here is to use the nested sampling algorithm, as implemented in the {\tt Multinest} software package~\cite{Feroz:2007kg,Feroz:2008xx,Feroz:2013hea}. This technique is particularly well suited to the problem but, to the best of our knowledge, has not been applied to composite Higgs models before now.

Our results comprise a set of overall tuning values as functions of phenomenologically interesting parameters: top partner masses, charged vector-boson resonance masses, deviations to the Higgs couplings, and the compositeness scale. We derive these for three explicit models. All are based on the minimal viable spontaneous symmetry breaking pattern in the composite Higgs sector, $SO(5)\to SO(4)$~\cite{hep-ph/0412089}, and on the simple, 4D constructions developed in refs.~\cite{DeCurtis:2011yx,1106.2719}. We consider three different choices for the $SO(5)$ embeddings of the top quark. The 5-5 model is the simplest, the 14-14 model is a slightly more involved model but allows for lower tuning, and the 14-1 model accommodates a fully-composite right-handed top quark.

From this data we find minimal values for the tuning compatible with current collider constraints on the top partner masses and deviations to the Higgs couplings, and we project how the tuning will worsen if no new physics is seen in the imminent future. We find values of 3.7\%, 5\% and 2.5\% for the current tuning in the 5-5, 14-14 and 14-1 models. Assuming a reach no greater than 2 TeV for top partner exclusion after 300 fb$^{-1}$ at the LHC~\cite{1311.0299,ColliderReach} these values decrease to around 1\%, 3.3\% and 0.8\% respectively. To beat the limits coming from top partner searches the Higgs couplings would need to be measured to an accuracy of a few percent. Charged vector-boson resonance searches have a much weaker effect on the tuning. In all three models the double tuning associated with simultaneously getting a light enough Higgs and the correct VEV is the main contribution.

The rest of this paper is organised as follows. In section~\ref{sec:tuning} we provide a general discussion on tuning in models predicting values for more than one observable, deriving a new tuning measure to properly account for multiple tunings. Section~\ref{sec:overview} contains an overview of the composite Higgs models we will be studying. Section~\ref{sec:scan} details our scanning procedure and our results are presented in section~\ref{sec:results}. We conclude in section~\ref{sec:conclusions}. The appendices give further information on the models studied.

\section{Tuning in more than one observable\label{sec:tuning}}

In any model predicting a value for more than one observable there is the possibility of multiple, independent occurrences of tuning. For the models we will study here values are predicted for three relevant observables: the Higgs VEV, the Higgs mass and the mass of the top quark. A tuning can be defined for each and the overall tuning in the model should be a combination of all three.\footnote{In a strict sense we do not actually predict the Higgs VEV as we rescale the compositeness scale to give the correct value. Given a full, UV complete theory the compositeness scale could not be freely scaled like this therefore any tuning in the ratio $\xi=v^2/f^2$ should still be taken into account.} To determine how to combine them one should first determine whether they are independent or not.

If all observed tuning stems from the Higgs VEV alone being highly sensitive to the input parameters then the tunings are not independent and the model only really suffers from a single tuning. The other two observables may, for example, be parameterised in terms of the Higgs VEV as $m_h=\lambda v$ and $m_t=y_tv$, so this kind of tuning occurs when the values of $\lambda$ and $y_t$ do not demonstrate a similar sensitivity. The other extreme scenario, triple tuning, occurs when all three values -- $v$, $\lambda$ and $y_t$ -- are highly sensitive to the input parameters. In other words one must tune the input parameters to get an acceptable value for the Higgs VEV, tune them again to get an acceptable value for the Higgs mass, then tune a third and final time to get an acceptable value for the mass of the top quark. Such a model is much more finely tuned and this should be reflected in the measure used to quantify the tuning.

The most widely used measure of a single tuning is the Barbieri-Giudice tuning~\cite{Barbieri:1987fn}
\be
\Delta_{\rm BG}=\norm{\pdiff{\ln{O}}{\ln{x}}}_{O=O_{\rm exp}}
\ee
which quantifies the tuning of the observable $O$, taking the experimentally measured value $O_{\rm exp}$, with respect to the input parameter $x$. In models with several input parameters, $\mathbf{x}$, it is often generalised to
\be
\Delta_{\rm BG}=\max\norm{\pdiff{\ln{O}}{\ln{x_i}}}_{O=O_{\rm exp}}.
\ee
One way of understanding this tuning measure is to think of the experimental result, $O=O_{\rm exp}$, as defining a surface in the parameter space. The Barbieri-Giudice tuning is then closely related to the magnitude of the normal of this surface~\cite{Casas:2004gh,Casas:2005ev}
\be
\Delta_{\rm BG}\sim\sqrt{\sum_i\norm{\pdiff{\ln{O}}{\ln{x_i}}}_{O=O_{\rm exp}}^2}=\norm{\nabla_lO}_{O=O_{\rm exp}}
\ee
where $[\nabla_l]_i\equiv\partial\ln/\partial\ln{x_i}$ is a gradient operator in the corresponding logarithmic parameter space, $\ln\mathbf{x}$. A large value results from a highly curved surface, matching our intuition about how a tuning measure ought to behave. Because of this we can think of the normals as defining `tuning vectors' for the model. Tuning vectors have the added advantage that their magnitudes are insensitive to the basis chosen for the parameter space so the tuning cannot be removed by making a clever choice of variables.

When trying to evaluate the overall tuning in a set of $n_O$ observables, $\mathbf{O}$, we start by writing down individual, single tunings for each observable using the tuning vectors above
\be
\Delta_1^a=\norm{\nabla_lO^a}_{\mathbf{O}=\mathbf{O}_{\rm exp}}
\ee
where $a$ runs over all observables and $\mathbf{O}=\mathbf{O}_{\rm exp}$ means that all observables are set to their experimentally measured values. The total single tuning is then given by the sum of the individual contributions normalised by the number of observables
\be
\Delta_1=\frac{1}{n_O}\sum_a\Delta_1^a.
\ee
In models that only suffer from tuning in a single observable the surfaces $O^a=O^a_{\rm exp}$ will all be approximately aligned and the tuning vectors approximately parallel. For example, the extreme case for the models studied here would be for the Higgs VEV to be the only value that depends on the input parameters at all. We would then find
\be
\nabla_lm_h=\nabla_lm_t=\nabla_lv
\ee
as $\partial\lambda/\partial x_i=\partial y_t/\partial x_i=0$, and the overall tuning would be given by
\be
\Delta_1=\frac{1}{3}\nbrack{\Delta_1^v+\Delta_1^{m_h}+\Delta_1^{m_t}}=\Delta_1^v.
\ee
This explains our choice of normalisation; we only want to count the unique source of tuning once in this limiting case.

When two of the tunings are independent it seems reasonable that the individual tunings defined above should be multiplied together somehow. A quantity that captures this behaviour is
\be\label{eq:Delta2ab}
\Delta_2^{ab}=\norm{\begin{array}{cc}
\nabla_lO^a.\nabla_lO^a & \nabla_lO^a.\nabla_lO^b \\
\nabla_lO^a.\nabla_lO^b & \nabla_lO^b.\nabla_lO^b \end{array}}^{1/2}_{\mathbf{O}=\mathbf{O}_{\rm exp}}.
\ee
For fully independent tunings the individual surfaces $O^a=O^a_{\rm exp}$ will not be aligned at all, the tuning vectors will be orthogonal, and $\nabla_lO^a.\nabla_lO^b$ will vanish to give $\Delta_2^{ab}=\Delta_1^a\Delta_1^b$. For fully dependent tunings, on the other hand, the tuning vectors are parallel so $\nabla_lO^a.\nabla_lO^b=\Delta_1^a\Delta_1^b$ and $\Delta_2^{ab}$ vanishes instead. To get an overall double tuning we evaluate this quantity for each possible pair of observables and take the sum. In the models studied here this gives
\be
\Delta_2=\frac{1}{2}\nbrack{\Delta_2^{(v,m_h)}+\Delta_2^{(v,m_t)}+\Delta_2^{(m_h,m_t)}}.
\ee
The normalisation is again chosen for the limiting, non-trivial case, where two tunings are independent and the third is fully dependent. Two terms in the sum will then be equal and the third will vanish to give $\Delta_2=\Delta_2^{(v,m_h)}$ (for example) so that we only count the unique source of double tuning once.

Finally, it is possible that all three tunings are independent and the model suffers from triple tuning. We can quantify this by extending the double tuning measure as defined in \eqref{eq:Delta2ab} in the obvious way
\be
\Delta_3^{abc}=\norm{\begin{array}{ccc}
\nabla_lO^a.\nabla_lO^a & \nabla_lO^a.\nabla_lO^b & \nabla_lO^a.\nabla_lO^c \\
\nabla_lO^a.\nabla_lO^b & \nabla_lO^b.\nabla_lO^b & \nabla_lO^b.\nabla_lO^c \\
\nabla_lO^a.\nabla_lO^c & \nabla_lO^b.\nabla_lO^c & \nabla_lO^c.\nabla_lO^c
\end{array}}^{1/2}_{\mathbf{O}=\mathbf{O}_{\rm exp}}.
\ee
This quantity again behaves as desired: it is zero if any two of the tuning vectors are parallel (i.e.\ the associated tunings are dependent) and evaluates to $\Delta_1^a\Delta_1^b\Delta_1^c$ if all three tuning vectors are orthogonal (i.e.\ the tunings are fully independent). Only one such quantity exists in the models studied here as there are only three observables being considered. More generally, one would sum over all distinct combinations of $a$, $b$ and $c$ and normalise accordingly.

Combining all three tuning measures gives the overall tuning
\be
\Delta=\Delta_1+\Delta_2+\Delta_3.
\ee
When significant triple tuning is present in a model it will dominate to give $\Delta\approx\Delta_3$. When significant double tuning is present but triple tuning is not one finds $\Delta\approx\Delta_2$. When only single tuning is present one finds $\Delta\approx\Delta_1$ and the tuning is closely related to the more commonly used Barbieri-Giudice measure.

Extending our measure to models predicting values for more observables is straightforward. One defines the $n_O\times n_O$ tuning matrix $[M_\Delta]^{ab}=\nabla_lO^a.\nabla_lO^b$. The $n$th tuning measure between a particular choice of $n$ observables is found by evaluating the determinant of the $n\times n$ submatrix containing only those rows and columns. The total $n$th tuning measure is the sum over all such submatrices normalised by $n_O-(n-1)$. This normalisation ensures that the one unique $n$th tuning is only counted once in the limiting case of exactly $n$ independent individual tunings.

Before moving on we point out that many properties of our tuning measure may well be captured by other measures in the literature~\cite{Athron:2007ry,Athron:2007qr}, particularly those based on a Bayesian approach~\cite{0705.0487,0812.0536,1204.4940,1208.0837,1312.4150}. Indeed, when there are exactly three observables the matrix appearing in our definition of triple tuning is the square of the Jacobian matrix encountered in a Bayesian approach. It seems quite possible that our tuning measure can be rigorously derived using a Bayesian approach with a suitable choice of priors for the complete set of parameters and experimental constraints for the complete set of observables (i.e.\ including the composite sector resonance masses and so on).

\section{Model overview\label{sec:overview}}

The most promising way for a composite Higgs to be realised is as a pseudo Nambu-Goldstone boson (pNGB) that is part of a hitherto unobserved composite sector, the composite sector itself emerging from a confining gauge theory (examples of suitable theories can be found in refs.~\cite{Miransky:1988xi,Galloway:2010bp,1211.7290,1311.6562,1312.5330,1312.5664,1402.0233,1404.7137,1502.07340,1506.00623}). Within this framework the minimal model compatible with custodial symmetry in the Higgs sector, and therefore precision electroweak measurements~\cite{hep-ph/0308036}, assumes a spontaneous symmetry breaking pattern $SO(5)\to SO(4)$ in the composite Higgs sector~\cite{hep-ph/0412089}. Four pNGBs are associated with this symmetry breaking pattern, exactly the right number to make up a Higgs doublet. Interactions of the pNGBs are determined by low energy theorems so all that is left is to couple the elementary, Standard Model (SM) fermions to the composite sector. This is invariably achieved using the idea of partial compositeness~\cite{Kaplan:1991dc,Gherghetta:2000qt}, whereupon the forms of the couplings are fixed once it is decided which $SO(5)$ representations the composite operators coupling to the elementary fermions are in or, equivalently, which representations the elementary fermions are embedded in.

Throughout this work we will focus on 4D models like those presented in ref.~\cite{DeCurtis:2011yx}, also utilising specific constructions found in refs.~\cite{Panico:2012uw,Carena:2014ria}. Expressions for all important quantities are taken directly from these references. These models provide an effective description of the pNGBs and the lowest-lying vector-boson and fermion resonances emerging from the composite sector. They have the advantage of being simple, calculable and, most importantly, having clear links with collider phenomenology. All of the models we will consider are based on the minimal symmetry breaking pattern $SO(5)\times U(1)_X\to SO(4)\times U(1)_X$ (the additional $U(1)_X$ is included to give the correct hypercharge for the fermions). Since these models have been comprehensively explained many times in the literature we will only give a brief overview here. Further details are provided in appendix~\ref{app:DMD}.

We will study three embeddings for the top quarks in detail. Embeddings for the lighter fermions are not generally specified as their contributions to the Higgs potential tend to be subdominant and will be neglected,\footnote{An exception in some models, that we do not study here, is the right-handed tau~\cite{1410.8555}.} although we will consider modifications to the $hb\bar{b}$ coupling as it is phenomenologically interesting. The three models and their $SO(5)\times U(1)_X$ embeddings for the left and right-handed top quarks, $q$ and $t^c$, are
\begin{itemize}
\item The 5-5 model with $q,t^c\in{\bf5}_{2/3}$~\cite{hep-ph/0612048,DeCurtis:2011yx}
\item The 14-14 model with $q,t^c\in{\bf14}_{2/3}$~\cite{Panico:2012uw,Carena:2014ria}
\item The 14-1 model with $q\in{\bf14}_{2/3}$ and $t^c\in{\bf1}_{2/3}$~\cite{Panico:2012uw,Carena:2014ria}
\end{itemize}
Our choice of embeddings is strongly motivated by ref.~\cite{Panico:2012uw}, where the authors show that these three models capture three qualitatively different scenarios for tuning in composite Higgs models. The 5-5 model relies on a cancellation between fermion contributions of different orders to generate the Higgs potential, the 14-14 model generates the potential using fermion contributions of a single order, and the 14-1 model works the same as the 14-14 model but also allows for a fully-composite right-handed top quark.

For each model we take the full set of composite sector input parameters -- a collection of masses, couplings and mixing parameters -- and use them to evaluate the Higgs VEV and the masses of the Higgs and top quark. In practise the Higgs VEV only appears in the ratio
\be
\xi\equiv\frac{v^2}{f^2}
\ee
where $f$ is the compositeness scale so we can simply rescale $f$ to give the correct Higgs VEV instead of treating $f$ as a separate input parameter. After doing this rescaling only points reproducing the observed values for the remaining two quantities are kept.

For each viable point we then evaluate the spectrum of resonances predicted. Of particular phenomenological importance are the masses and SM quantum numbers of the lightest top partners. These are coloured, vector-like fermions that mix with the top quark and enable it to couple to the composite Higgs. They are vital for generating both a large top quark Yukawa coupling and a viable Higgs potential~\cite{Matsedonskyi:2012ym,Marzocca:2012zn,Pomarol:2012qf,Panico:2012uw,Pappadopulo:2013vca,Barnard:2013hka,1405.1498}, and usually provide the best collider signal.\footnote{Similar states exist for the other SM fermions but, since the other SM fermions couple much less strongly to the Higgs, they have a much smaller effect on the Higgs potential and their masses are not as well constrained by such considerations.} Also important are the lightest vector-boson resonances charged under the electroweak gauge group, although these provide weaker and more model-dependent constraints than the top partners.

Lastly we evaluate any deviations of the Higgs couplings to other SM fields relative to their SM values. These are parameterised using the ratios
\be
r_\chi\equiv\frac{c(h\chi\chi)}{c_{\rm SM}(h\chi\chi)}
\ee
for couplings $c$ and $c_{\rm SM}$ in the composite Higgs model and SM respectively.

\subsection{Universal features}

Several aspects of the models we are interested in, particularly those concerning the gauge sector, only depend on the global symmetry breaking pattern so are equivalent for all choices of matter embedding.

The gauge sector is described by an angle, $t_\theta\equiv\tan\theta$, quantifying the amount of elementary-composite mixing in the gauge sector and taken to be small; a mass, $m_\rho$ for the lightest vector-boson resonance; and a mass, $m_a$ of another, heavier vector-boson resonance. We allow these parameters to freely vary in the intervals
\begin{align}
t_\theta & \in[0,1] &
m_\rho, m_a & \in[0.5,10]\mbox{ TeV}
\end{align}
with $m_a>m_\rho$. Here and in the fermion sector we check that the value of $f$ found for each point is consistent with all dimensionful parameters having magnitudes less than $4\pi f$.

The spectrum of massive vector-boson resonances coming from the composite sector includes several states charged under $SU(2)_L\times U(1)_Y$. The quantum numbers and masses (up to small, post electroweak symmetry breaking corrections) of the lightest charged states are given by
\begin{itemize}
\item ${\bf1}_{\pm1}$ with mass $m_{\rho_{\bf1}}=m_\rho$
\item ${\bf3}_0$ with mass $m_{\rho_{\bf3}}=m_\rho/c_\theta$
\end{itemize}
where $c_\theta\equiv\cos\theta$.

The modification to the $hVV$ coupling, where $V$ is a $W$ or $Z$ boson, is given by
\be
r_V=\sqrt{1-\xi}.
\ee
Modifications to the $htt$ and $hbb$ couplings depend on the embeddings chosen for the quarks and will be described in the following subsections. Up to small contributions from lighter states the modification to the loop-induced $hgg$ coupling, $r_g$, is the same as that of the $htt$ coupling. The modification to the loop-induced $h\gamma\gamma$ coupling is determined from the previous quantities via the formula
\be
r_\gamma=\frac{A_1r_V+\frac{4}{3}A_{1/2}r_t}{A_1+\frac{4}{3}A_{1/2}}
\ee
neglecting contributions from states lighter than the $W$ and $Z$ bosons. The $A$'s are the standard $W$ and $Z$ boson and top quark loop functions, evaluating to $A_1\approx-8.324$ and $A_{1/2}\approx1.375$ respectively.

\subsection{5-5 model overview}

The simplest model embeds both the left and right-handed top quarks in the ${\bf5}_{2/3}$ representation of $SO(5)\times U(1)_X$. The parameters of the composite sector comprise a Yukawa-like coupling, $Y$, coupling the top partners to the Higgs; top partner mass terms $m_Q$, $m_T$ and $m_Y$; and elementary-composite mixing parameters, $d_q$ and $d_t$. We allow the parameters to freely vary in the intervals
\begin{align}
m_Q, m_T, m_Y & \in[0.5,10]\mbox{ TeV} &
Y & \in[-10,10]\mbox{ TeV} &
d_q, d_t & \in[0,1].
\end{align}

There are two top partners, also in the ${\bf5}_{2/3}$ representation. Breaking these down into $SU(2)_L\times U(1)_Y$ multiplets the lightest states are
\begin{itemize}
\item ${\bf1}_{2/3}=T_{2/3}$ with mass $m_{{\bf1}_{2/3}}$
\item ${\bf2}_{1/6}=(T_{2/3},\, B_{-1/3})$ with mass $m_{{\bf2}_{1/6}}$
\item ${\bf2}_{7/6}=(T_{5/3},\, B_{2/3})$ with mass $m_{{\bf2}_{7/6}}$
\end{itemize}
The remaining modifications to the Higgs couplings are given by
\begin{align}
r_g=r_t=r_b & =\frac{1-2\xi}{\sqrt{1-\xi}}.
\end{align}

In this model the Higgs quartic coupling is generated at a higher order than the quadratic coupling (specifically at quartic rather than quadratic order in the elementary-composite fermion mixing parameters). A high degree of tuning is therefore expected. Further details for this model are given in appendix~\ref{app:55}.

\subsection{14-14 model overview}

This model embeds both the left and right-handed top quarks in the ${\bf14}_{2/3}$ representation of $SO(5)\times U(1)_X$. The parameters of the composite sector comprise two Yukawa-like couplings, $Y_1$ and $Y_2$, coupling the top partners to the Higgs; top partner mass terms $m_Q$, $m_T$ and $m_Y$; and elementary-composite mixing parameters, $d_q$ and $d_t$. We allow the parameters to freely vary in the intervals
\begin{align}
m_Q, m_T, m_Y & \in[0.5,10]\mbox{ TeV} &
Y_1,Y_2 & \in[-10,10]\mbox{ TeV} &
d_q, d_t & \in[0,1].
\end{align}

There are two top partners, also in the ${\bf14}_{2/3}$ representation. Breaking these down into $SU(2)_L\times U(1)_Y$ multiplets the lightest states are
\begin{itemize}
\item ${\bf1}_{2/3}=T_{2/3}$ with mass $m_{{\bf1}_{2/3}}$
\item ${\bf2}_{1/6}=(T_{2/3},\, B_{-1/3})$ with mass $m_{{\bf2}_{1/6}}$
\item ${\bf2}_{7/6}=(T_{5/3},\, B_{2/3})$ with mass $m_{{\bf2}_{7/6}}$
\item ${\bf3}_{5/3}+{\bf3}_{2/3}+{\bf3}_{-1/3}$ with mass $m_{\bf3}$
\end{itemize}
The remaining modifications to the Higgs couplings are given by
\begin{align}
r_g=r_t & =\frac{5(1-8\xi+8\xi^2)Y_1-2(4-23\xi+20\xi^2)Y_2}{2\xi(1-\xi)[5(2\xi-1)Y_1+2(4-5\xi)Y_2]} & r_b & =\frac{1-2\xi}{\sqrt{1-\xi}}.
\end{align}

Unlike the 5-5 model, the entire Higgs potential is generated at the same order (quadratic order in the elementary-composite fermion mixing parameters). Hence the tuning can be expected to be less severe. Further details for this model are given in appendix~\ref{app:1414}.

\subsection{14-1 model overview}

This model embeds the left-handed top quark in the ${\bf14}_{2/3}$ representation of $SO(5)\times U(1)_X$ and the right-handed top quark in the ${\bf1}_{2/3}$ representation. The parameters of the composite sector comprise two Yukawa-like couplings, $Y_1$ and $Y_2$, coupling the top partners to the Higgs; top partner mass terms $m_{Q_1}$, $m_{Q_2}$ and $m_Y$; and elementary-composite mixing parameters, $d_q$ and $\Lambda$. We allow the parameters to freely vary in the intervals
\begin{align}
m_{Q_1}, m_{Q_2}, m_Y,\Lambda & \in[0.5,10]\mbox{ TeV} &
Y_1, Y_2 & \in[-10,10]\mbox{ TeV} &
d_q & \in[0,1].
\end{align}

There are two top partners, also in the ${\bf14}_{2/3}$ representation. The second top partner is needed to break an accidental symmetry that would otherwise increase the tuning. Breaking these down into $SU(2)_L\times U(1)_Y$ multiplets the lightest states are the same as in the 14-14 model, i.e.\
\begin{itemize}
\item ${\bf1}_{2/3}=T_{2/3}$ with mass $m_{{\bf1}_{2/3}}$
\item ${\bf2}_{1/6}=(T_{2/3},\, B_{-1/3})$ with mass $m_{{\bf2}_{1/6}}$
\item ${\bf2}_{7/6}=(T_{5/3},\, B_{2/3})$ with mass $m_{{\bf2}_{7/6}}$
\item ${\bf3}_{5/3}+{\bf3}_{2/3}+{\bf3}_{-1/3}$ with mass $m_{\bf3}$
\end{itemize}
The remaining modifications to the Higgs couplings are the same as in the 5-5- model and are given by
\begin{align}
r_g=r_t=r_b & =\frac{1-2\xi}{\sqrt{1-\xi}}.
\end{align}

Once again the entire Higgs potential is generated at the same order (quadratic order in the elementary-composite fermion mixing parameters) so low tuning could be expected. Embedding the right-handed top quark in the ${\bf1}$ of $SO(5)$ also allows for it to be fully composite, as there is no need for its couplings to explicitly break $SO(5)$. Further details for this model are given in appendix~\ref{app:141}.

\section{Scanning procedure\label{sec:scan}}

Given a set of input parameters, $\mathbf{x}$, we wish to obtain the region of the parameter space in which the masses predicted for the Higgs and the top quark closely match the observed values. As the expressions involved cannot easily be algebraically inverted the simplest approach would be to scan over each parameter using a flat grid scan. Unfortunately this scales badly with the dimension of the parameter space, which is between nine and ten for the models studied here. On the other hand the nested sampling algorithm, as implemented in the {\tt Multinest} software package, has proven particularly useful for sampling the non-trivial (and possibly multimodal) functions encountered in many particle physics and cosmology examples. We here briefly summarise the application of the technique to our particular problem leaving the finer details of nested sampling itself to the original papers~\cite{Feroz:2007kg,Feroz:2008xx,Feroz:2013hea}.

Given $\mathbf{O}\equiv \{m_h,m_t\}$ the likelihood of any particular model with $N_p$ parameters $\mathbf{x}$ is
\be
p(\mathbf{O}|\mathbf{x})=\prod_a\exp\nbrack{-\frac{[O^a(\mathbf{x})-O_{\rm exp}^a]^2}{2(\sigma^a)^2}}
\ee
where $O^a(\mathbf{x})$ is the predicted value of the $i$th observable with experimentally measured value $O^a_{\rm exp}$, $\sigma^a$ is the error in $O^a_{\rm exp}$, and the product runs over all observables. For our purposes $\sigma^a$ characterises how close we want the masses to be to their observed values. Given a prior knowledge, $p(\mathbf{x})$, of the distribution of model parameters we can determine the posterior probability of $\mathbf{x}$ via Bayes' theorem
\be
p(\mathbf{x}|\mathbf{O}) = \frac{p(\mathbf{O}|\mathbf{x})p(\mathbf{x})}{Z}.
\ee
The points giving higher posterior probabilities predict more viable top quark and Higgs masses.

The normalisation constant, $Z$, is the Bayesian evidence
\be\label{evidence}
Z=\int p(\mathbf{O}|\mathbf{x})(\mathbf{x}) p(\mathbf{x})d^{N_{\rm p}}\mathbf{x}.
\ee
Nested sampling is a Monte Carlo method that calculates the evidence by transforming this difficult, multi-dimensional evidence integral into a one-dimensional integral that is easy to evaluate numerically. As a by-product one obtains posterior samples and it is these sample points that we will interpret in the following sections.

We use a flat prior on all parameters in this study. The choice of prior ultimately determines the sampling density in the final regions of interest but does not have a dominant effect on whether those regions are found or not. Given that the object of this study is to locate viable regions of parameter space, irrespective of the final sampling density, a study with one prior is sufficient. We have checked that doubling the number of live points in the {\tt Multinest} algorithm from 4000 to 8000 gives similar results in terms of the located regions, suggesting that we have good coverage of the parameter space.

For the Higgs and top quark masses we choose central values of $125$ GeV and $155$ GeV, and `errors' of $5$ GeV and $15$ GeV. These choices result in points with $m_h\in[120,130]$ GeV and $m_t\in[140,170]$ GeV being favoured. The low value and large interval for the top quark mass is chosen to account for running down to the electroweak scale and converting to the pole mass, the original prediction giving the running mass at a variable scale $f\sim$ few TeV\@. Recall that the Higgs VEV only appears in the ratio $\xi\equiv v^2/f^2$ so no additional input is required; we can simply rescale $f$ to give the correct Higgs VEV\@.

To account for collider constraints we assign a likelihood of zero to any point where at least one of the top partner masses does not satisfy the CMS limits $m_{{\bf2}_{1/6}}>786$ GeV, $m_{{\bf1}_{2/3}}>696$ GeV, $m_{{\bf2}_{7/6}}>800$ GeV and $m_{\bf3}>800$ GeV \cite{Chatrchyan:2013uxa,Chatrchyan:2013wfa,CMS:2013una} (see also refs.~\cite{1401.2457,1502.04718,1506.01961}). Note that these limits only apply to a single top partner in isolation. Since the models we consider all contain more than one top partner the actual limits will be stronger. A full reinterpretation of the analyses would be required to quantify this and is beyond the scope of this work. Limits on the vector-boson resonance masses are not applied. In a simplified approach the current limits are around $1.5$ TeV~\cite{Khachatryan:2014xja} but they are often weakened by model dependent effects. We anyway find that these masses are not so strongly correlated with the tuning so such limits are unlikely to have a significant impact on our results. Limits on these models from Higgs coupling measurements have been derived in ref.~\cite{Falkowski:2013dza} and imply that $f\gtrsim700$ GeV or, equivalently, $\xi\lesssim0.12$. Constraints from electroweak precision measurements and flavour-changing effects are all model dependent and are not applied here. Our results therefore provide a conservative lower bound on the tuning.\footnote{Additional couplings can be included that only really influence electroweak precision observables and the flavour sector. The extra constraints will generally impose limits, and perhaps tunings, on this extended parameter space while leaving the minimal parameter space largely unaffected. A caveat to this statement comes from the lepton sector. Increased levels of tau compositeness, motivated by flavour considerations, can actually lessen the dependence of the Higgs mass on the top partner masses and reduce the tuning~\cite{1410.8555}.}

\section{Results\label{sec:results}}

\begin{figure}
\begin{center}
\includegraphics[width=0.47\textwidth]{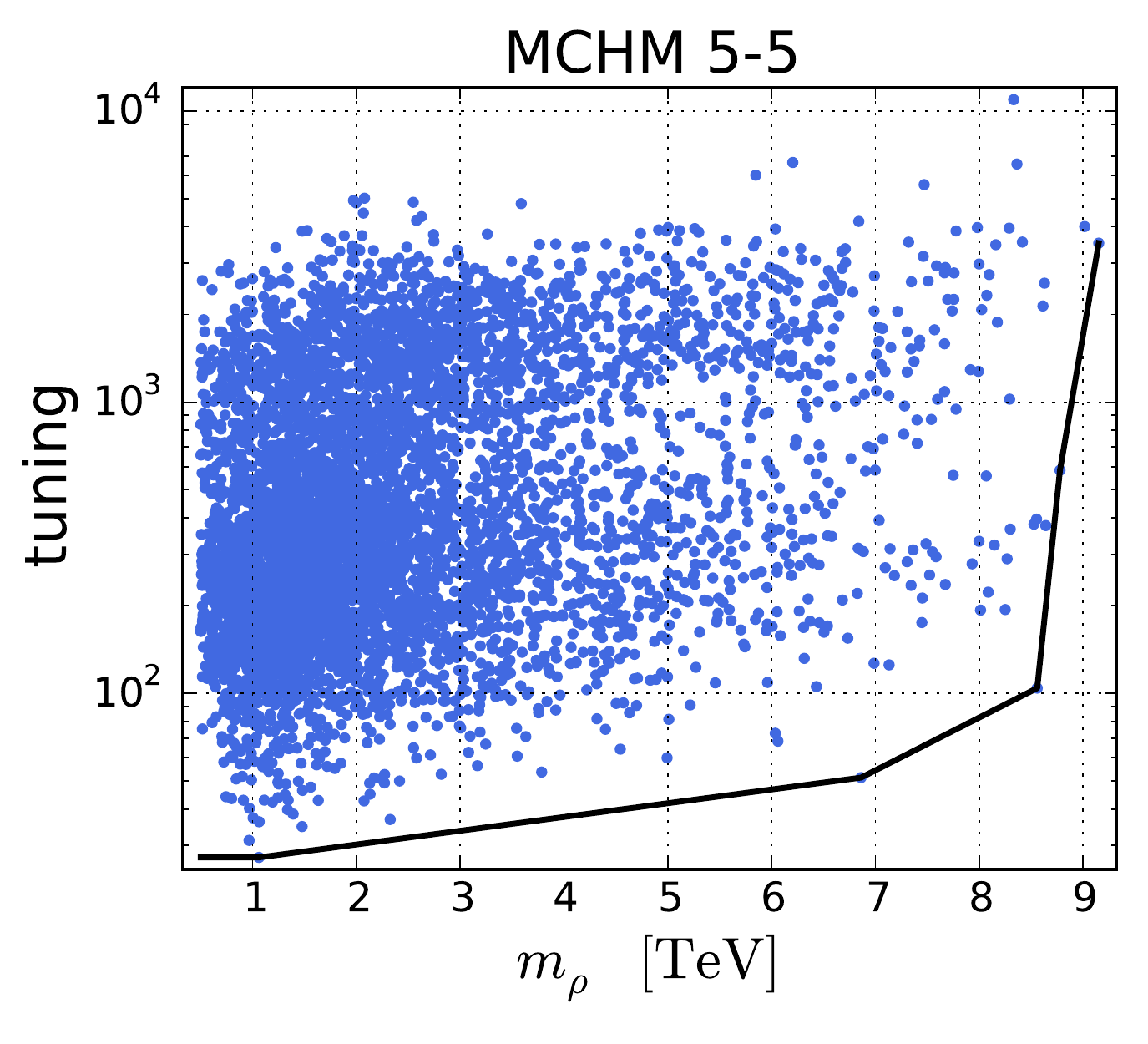}\hspace{5mm}
\includegraphics[width=0.47\textwidth]{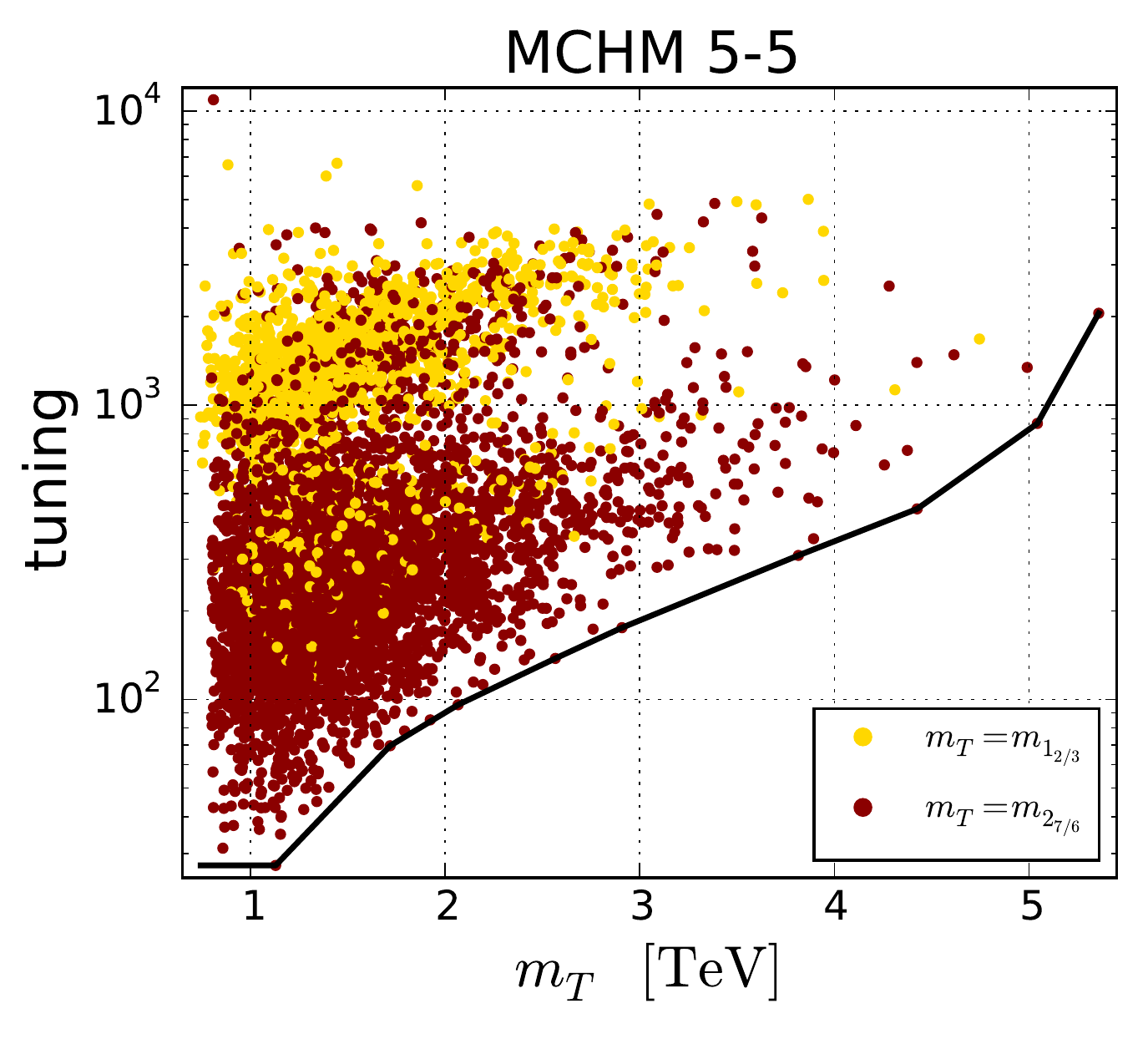}\\
\includegraphics[width=0.47\textwidth]{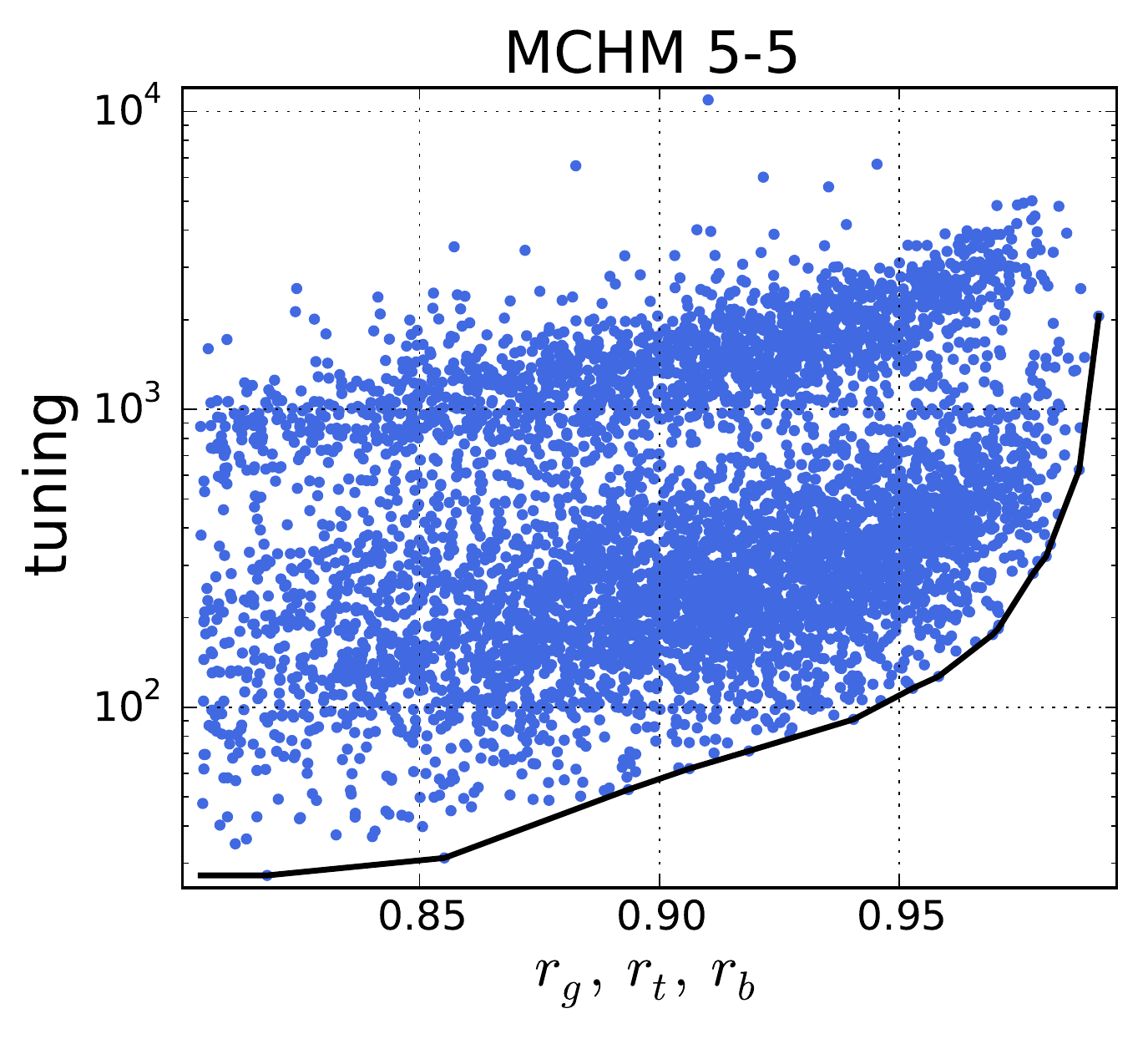}\hspace{5mm}
\includegraphics[width=0.47\textwidth]{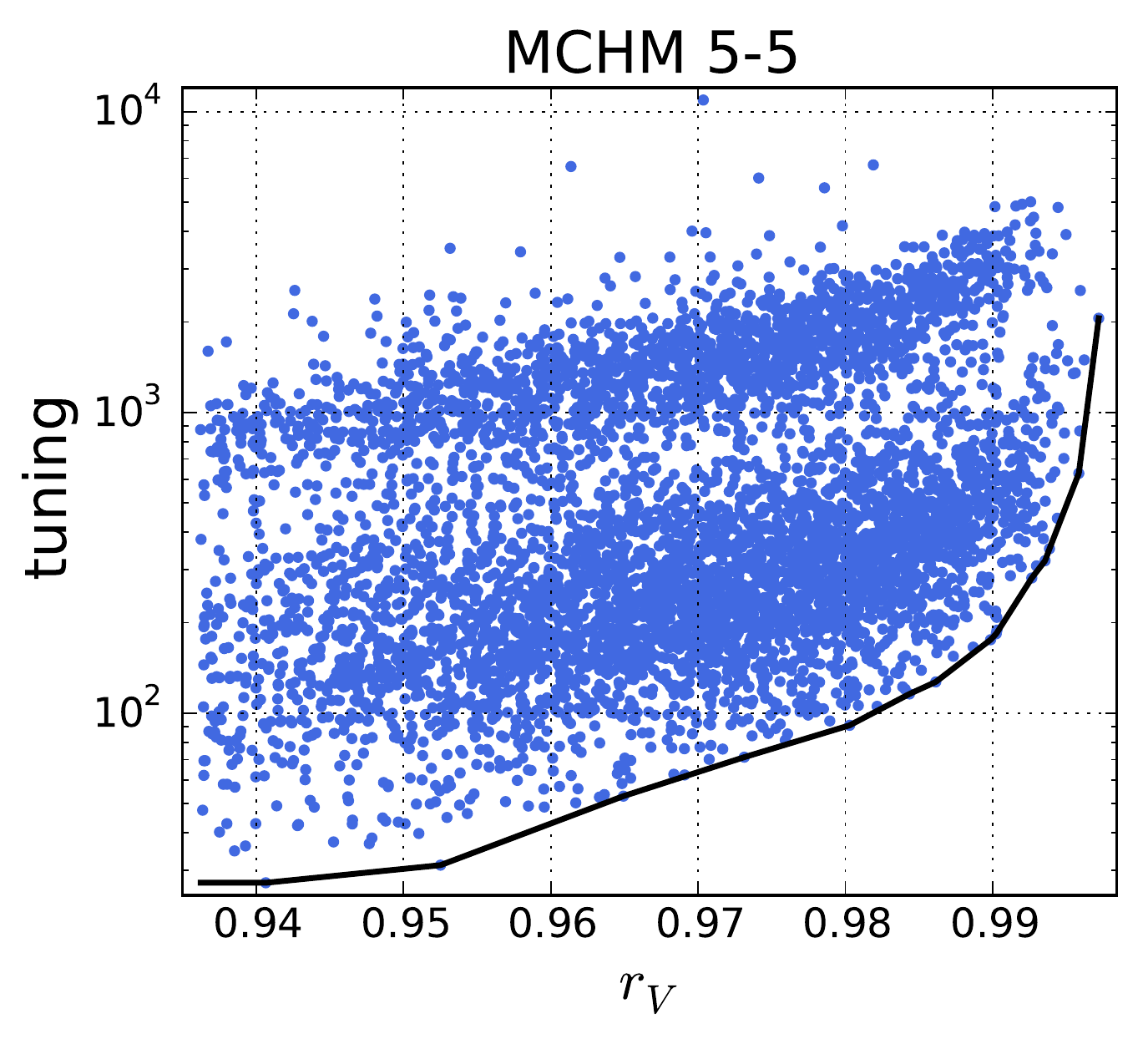}\\
\includegraphics[width=0.47\textwidth]{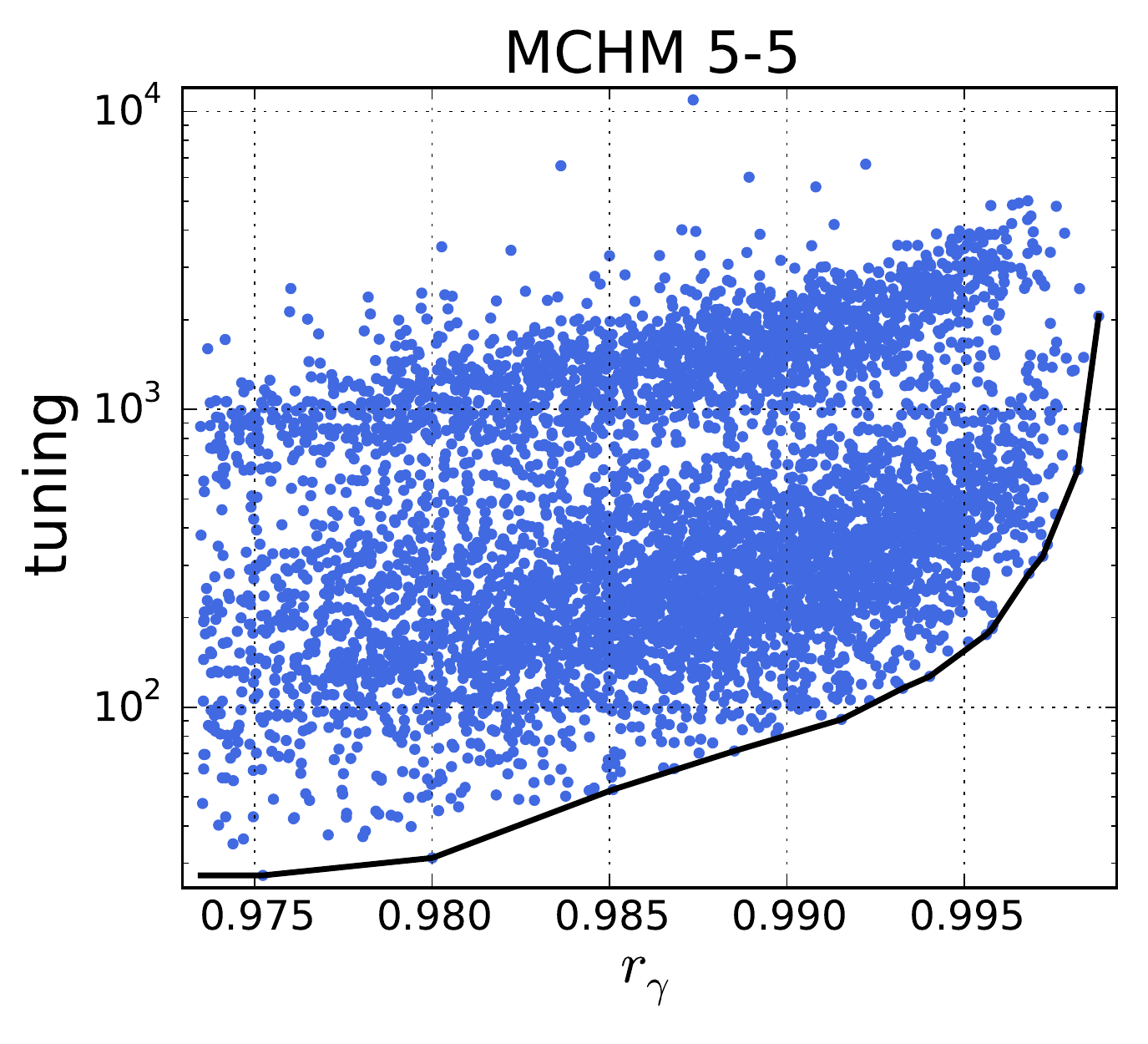}\hspace{5mm}
\includegraphics[width=0.47\textwidth]{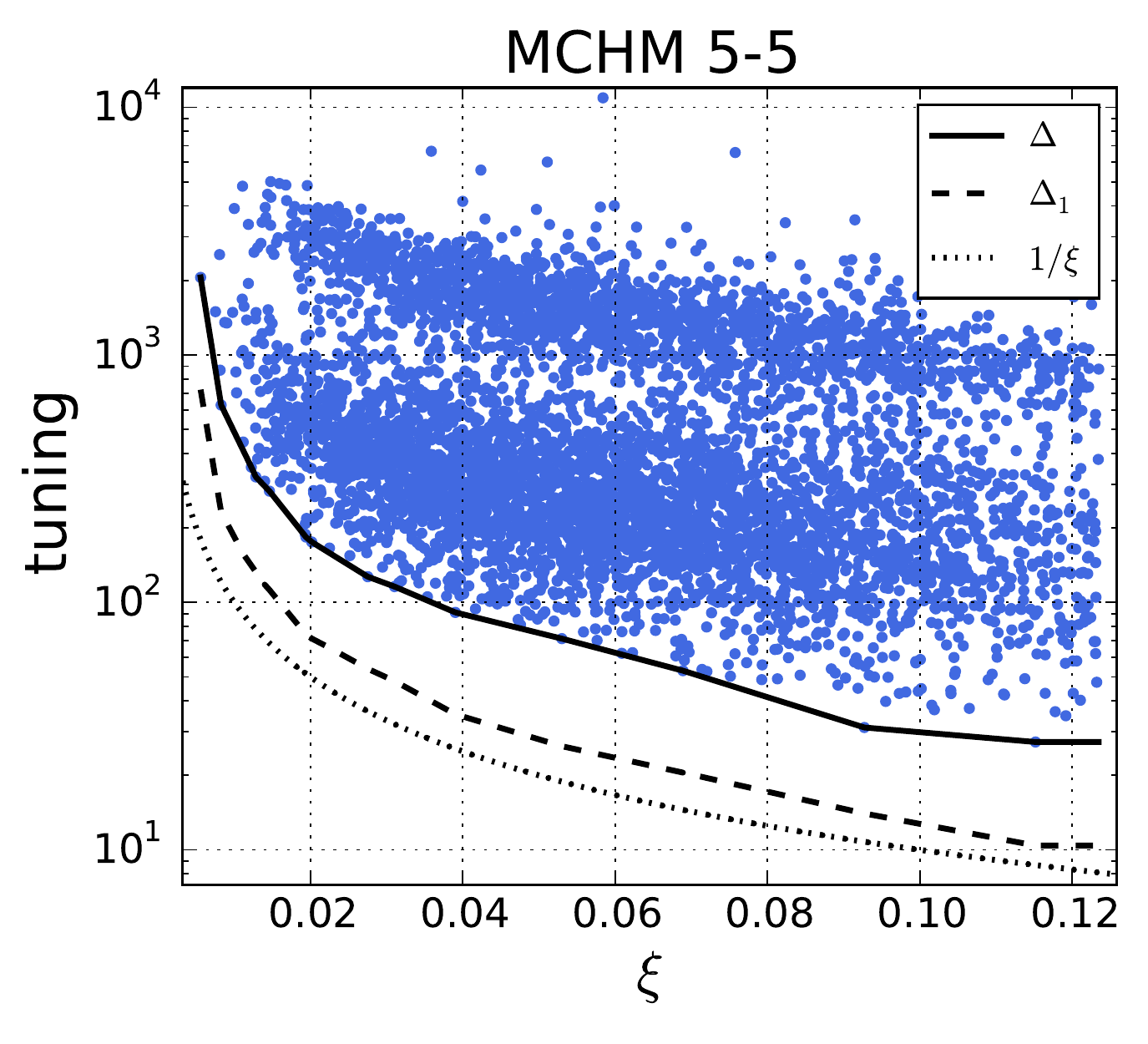}\vspace{-5mm}
\end{center}
\caption{Tuning in the 5-5 model as a function of $\rho$ mass, top partner mass, Higgs couplings and $\xi$. Further details are given in the text.\label{fig:55r}}
\end{figure}

After performing the scan detailed above we evaluated the tuning for each point as detailed in section~\ref{sec:tuning}. Tunings for the three models are then plotted against the mass, $m_\rho$, of the lightest vector-boson resonance, which has SM quantum numbers ${\bf1}_{\pm1}$; the mass, $m_T$, of the lightest top partner resonance, which has different SM quantum numbers detailed by colour coding of the plots; the ratios, $r_\chi$, of the Higgs couplings to their SM values, and the ratio, $\xi\equiv v^2/f^2$, of the Higgs VEV to the compositeness scale. These plots are shown in figures~\ref{fig:55r}, \ref{fig:1414r} and \ref{fig:141r}. In all plots we provide a simple lower bound on the tuning by connecting the extremal points using a convex hull.

Double tuning dominates in all models, specifically a double tuning between the Higgs VEV and mass. This can be seen in the plots of tuning against $\xi$ where the lower bound from the single tuning, $\Delta_1$, is shown alongside the overall tuning (we also include the naive tuning measure, $1/\xi$, in these plots for comparison). We can understand this from the arguments given in refs.~\cite{Matsedonskyi:2012ym,Marzocca:2012zn,Pomarol:2012qf,Panico:2012uw,Pappadopulo:2013vca,Barnard:2013hka}. A light Higgs can only be obtained without double tuning in the presence of light top partners but the CMS limits are already disfavouring this possibility. Triple tuning was not significant in any of the models we studied.

\subsection{5-5 model results}

Full results for the 5-5 model are shown in figure~\ref{fig:55r}. The minimum tuning we find is $\Delta=27$, i.e.\ about 3.7\%. Previous estimates of the tuning in this model have been a little less severe as they did not consider the double tuning required to get a light enough Higgs. This is shown explicitly in figure~\ref{fig:55r}, where the single tuning is shown to be consistently below the overall tuning by a factor of two to three. The naive, $1/\xi$ estimate of tuning is a little smaller still.

Points with milder tuning predict lighter top partners in the ${\bf2}_{7/6}$ SM multiplet, a lower compositeness scale, and a high degree of elementary-composite mixing in the top quark sector. This all leads to larger modifications to the couplings of the Higgs to gluons and fermions, and the tuning can be well constrained in this model by measuring these couplings more precisely. Once a precision better than 5\% is achieved the model very quickly moves beyond $\Delta=100$ and begins to look unnatural. Further constraints on the top partner masses also quickly lead to a higher degree of tuning. With a final top partner reach of about 2 TeV the LHC can push the tuning to $\Delta\gtrsim100$ from these searches alone.

Points giving the correct Higgs mass in this model clearly separate into two regions: one in which the ${\bf2}_{7/6}$ SM multiplet tends to be the lightest top partner and one in which it tends to be the ${\bf1}_{2/3}$. The latter region is significantly more tuned. This is because it is harder to keep the ${\bf1}_{2/3}$ light than the ${\bf2}_{7/6}$. The ${\bf1}_{2/3}$ mixes directly with the elementary, right-handed top quark so its mass is more constrained by the observed top quark mass. The ${\bf2}_{7/6}$ does not mix directly with the elementary top quark so its mass is more flexible.

\subsection{14-14 model results}

\begin{figure}
\begin{center}
\includegraphics[width=0.47\textwidth]{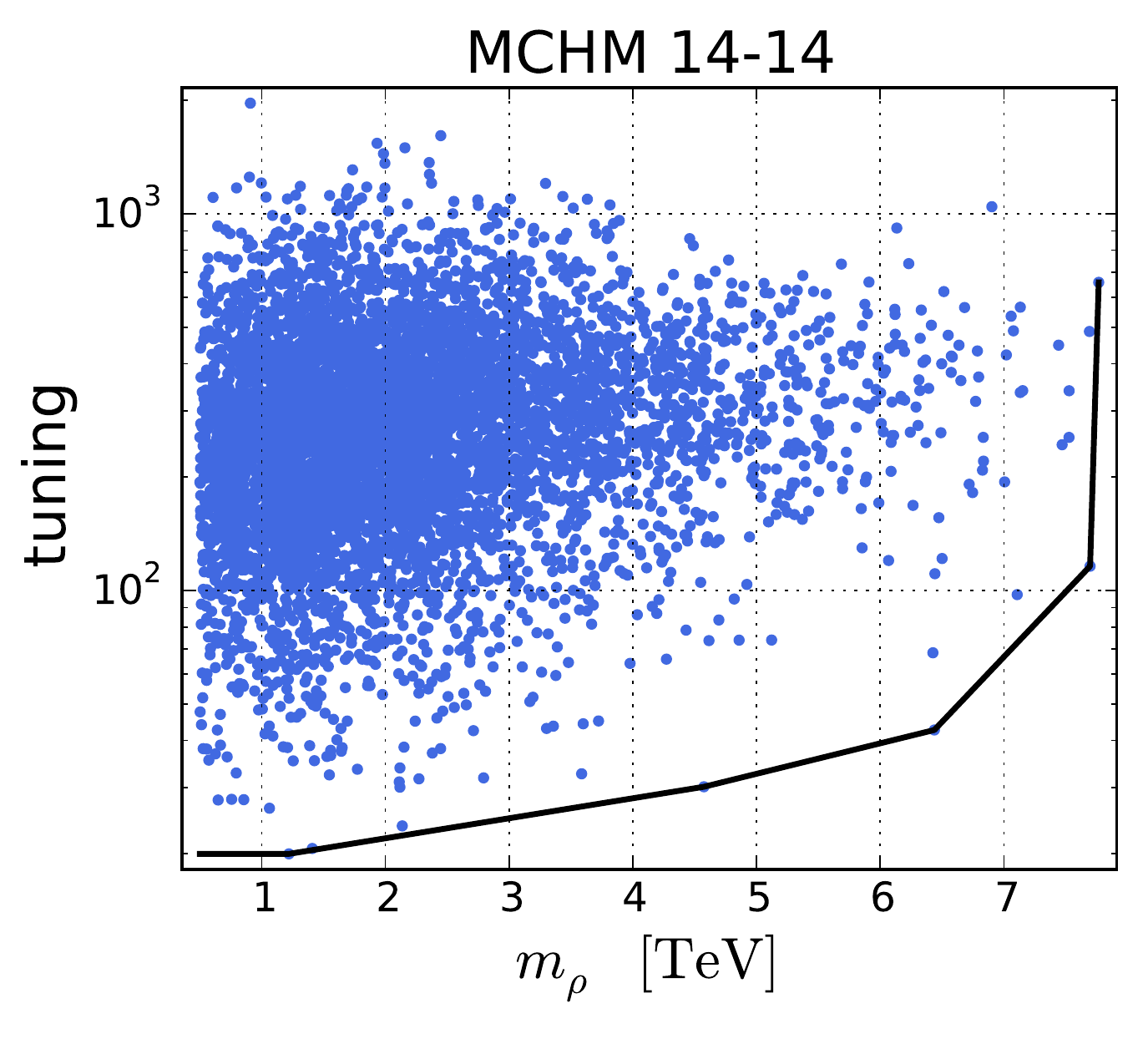}\hspace{5mm}
\includegraphics[width=0.47\textwidth]{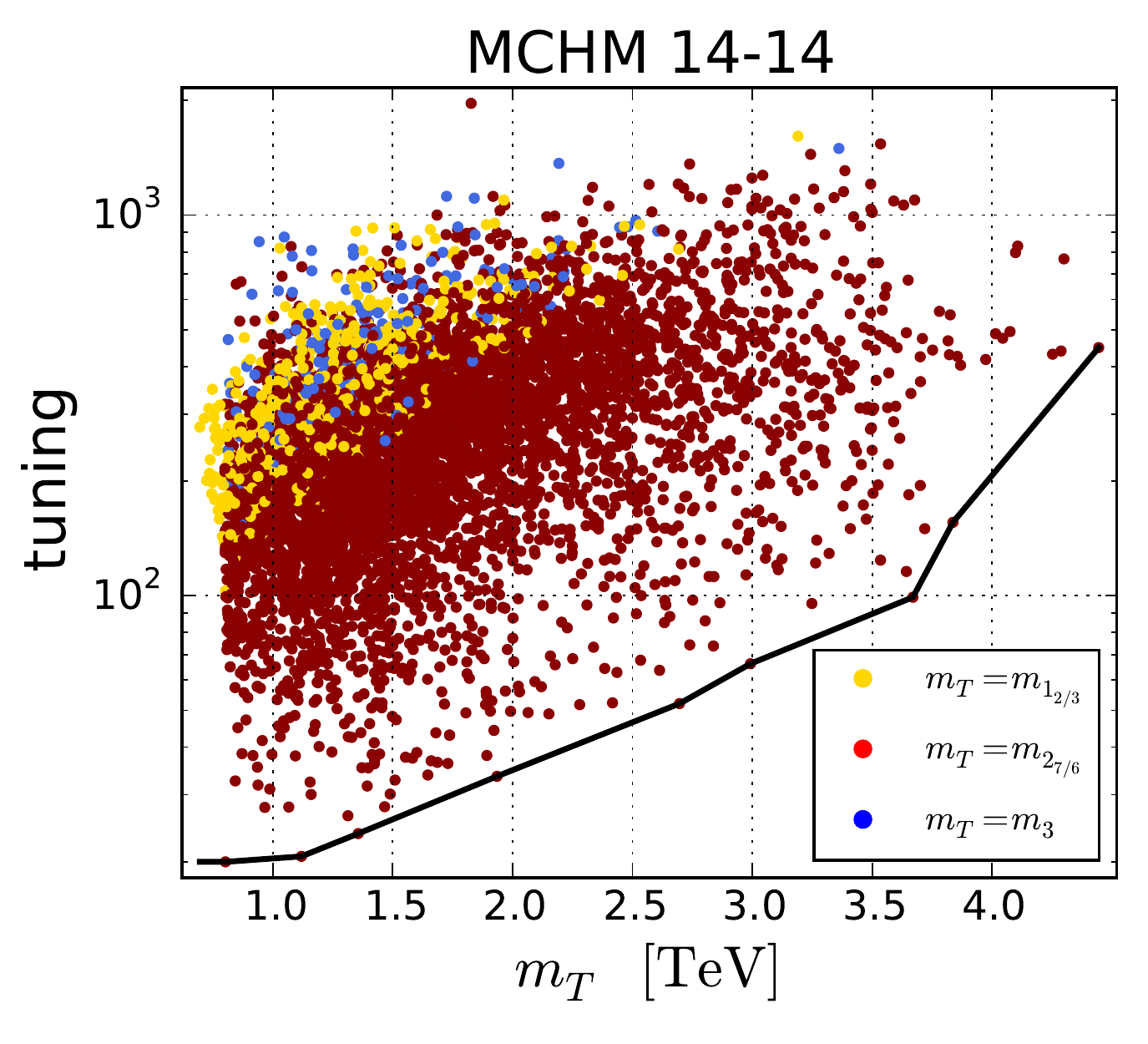}\\
\includegraphics[width=0.47\textwidth]{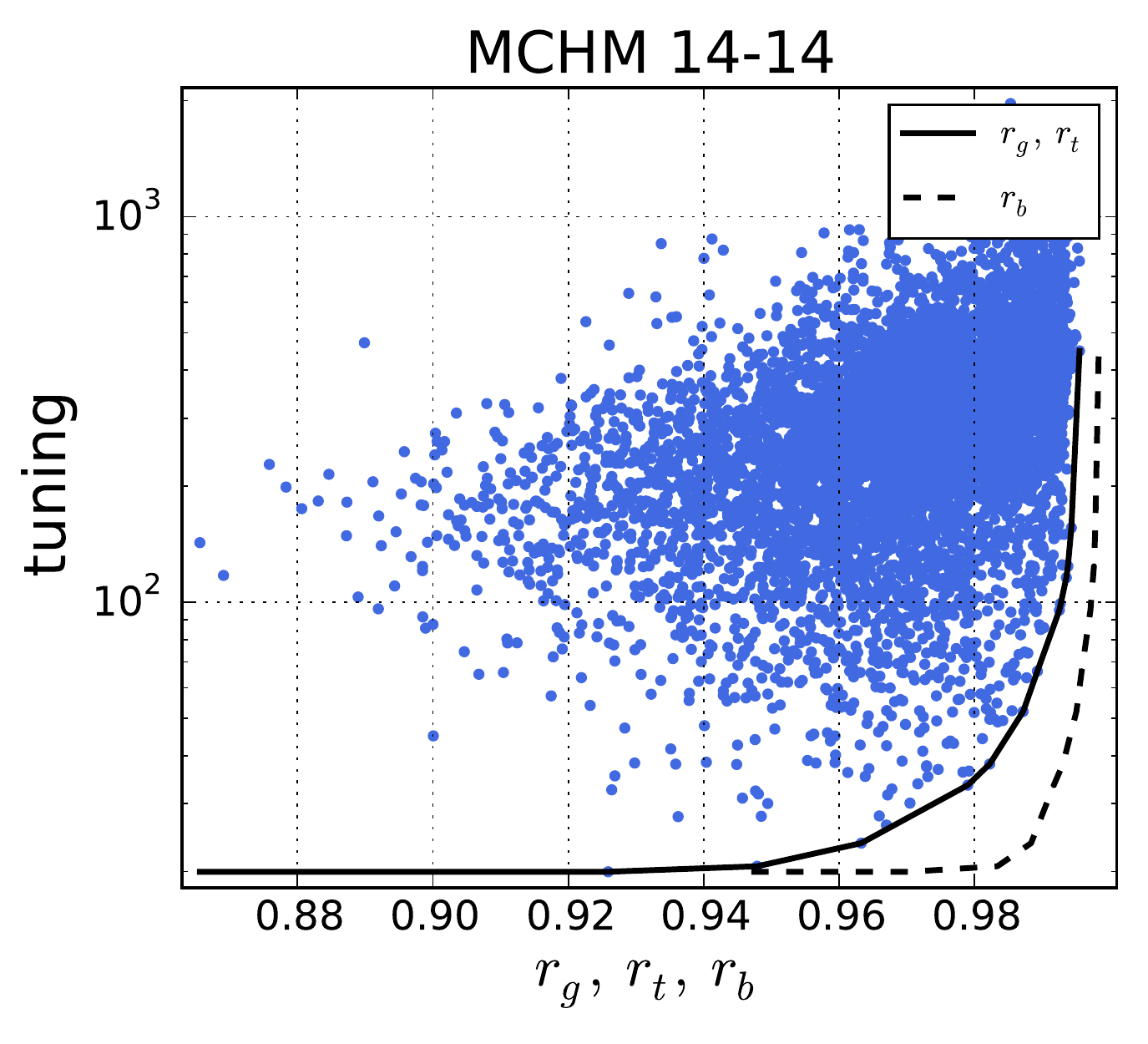}\hspace{5mm}
\includegraphics[width=0.47\textwidth]{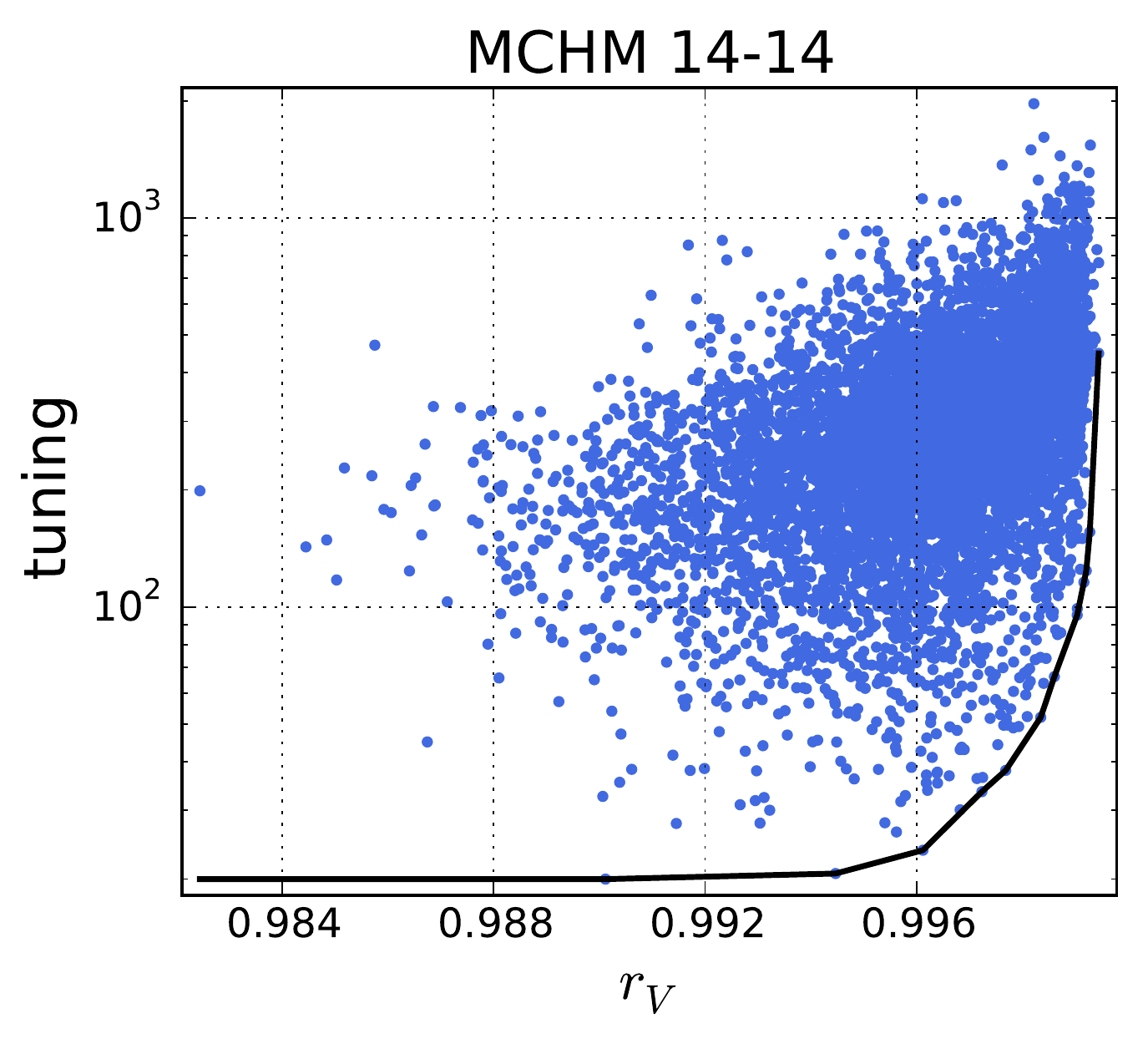}\\
\includegraphics[width=0.47\textwidth]{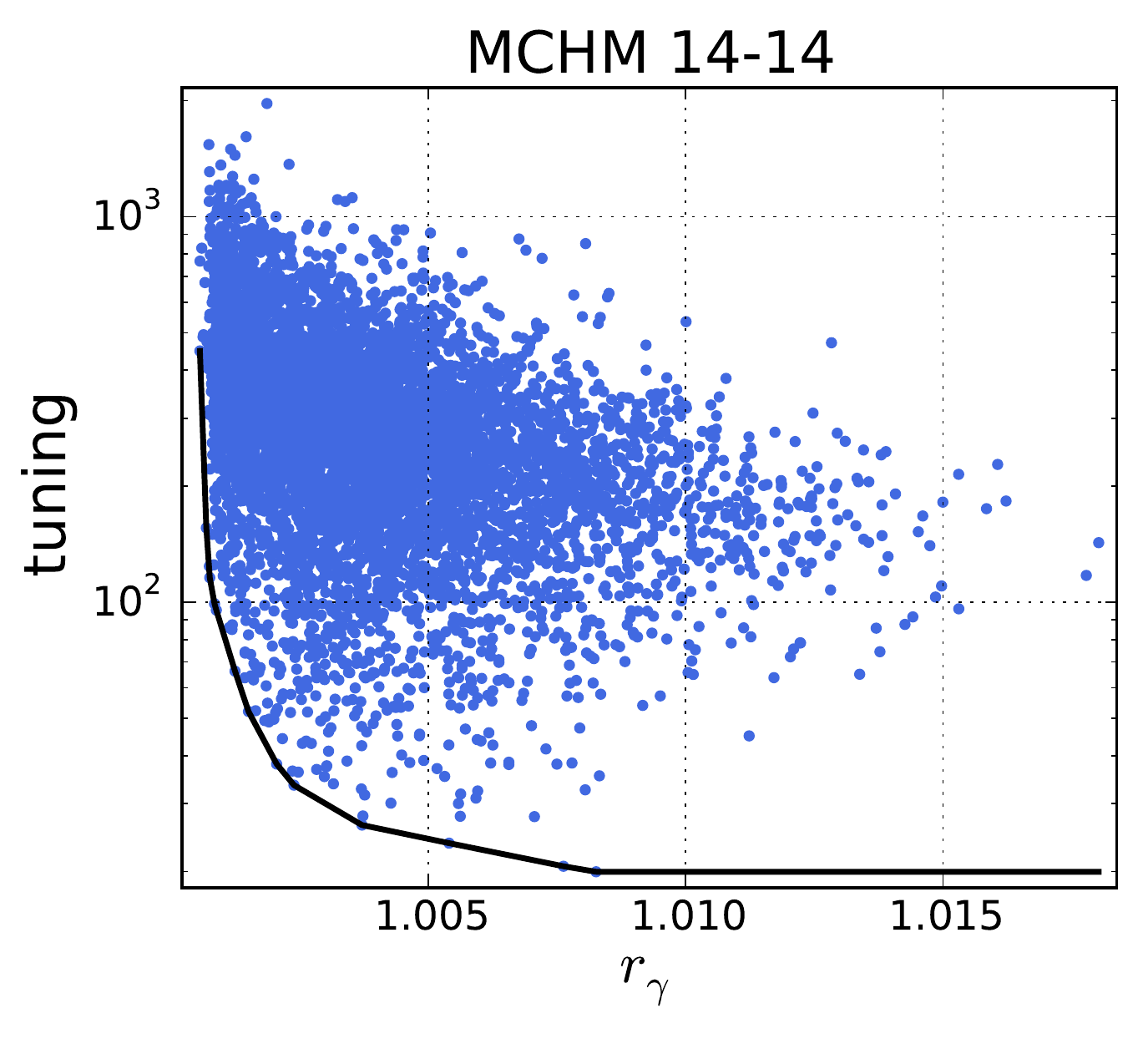}\hspace{5mm}
\includegraphics[width=0.47\textwidth]{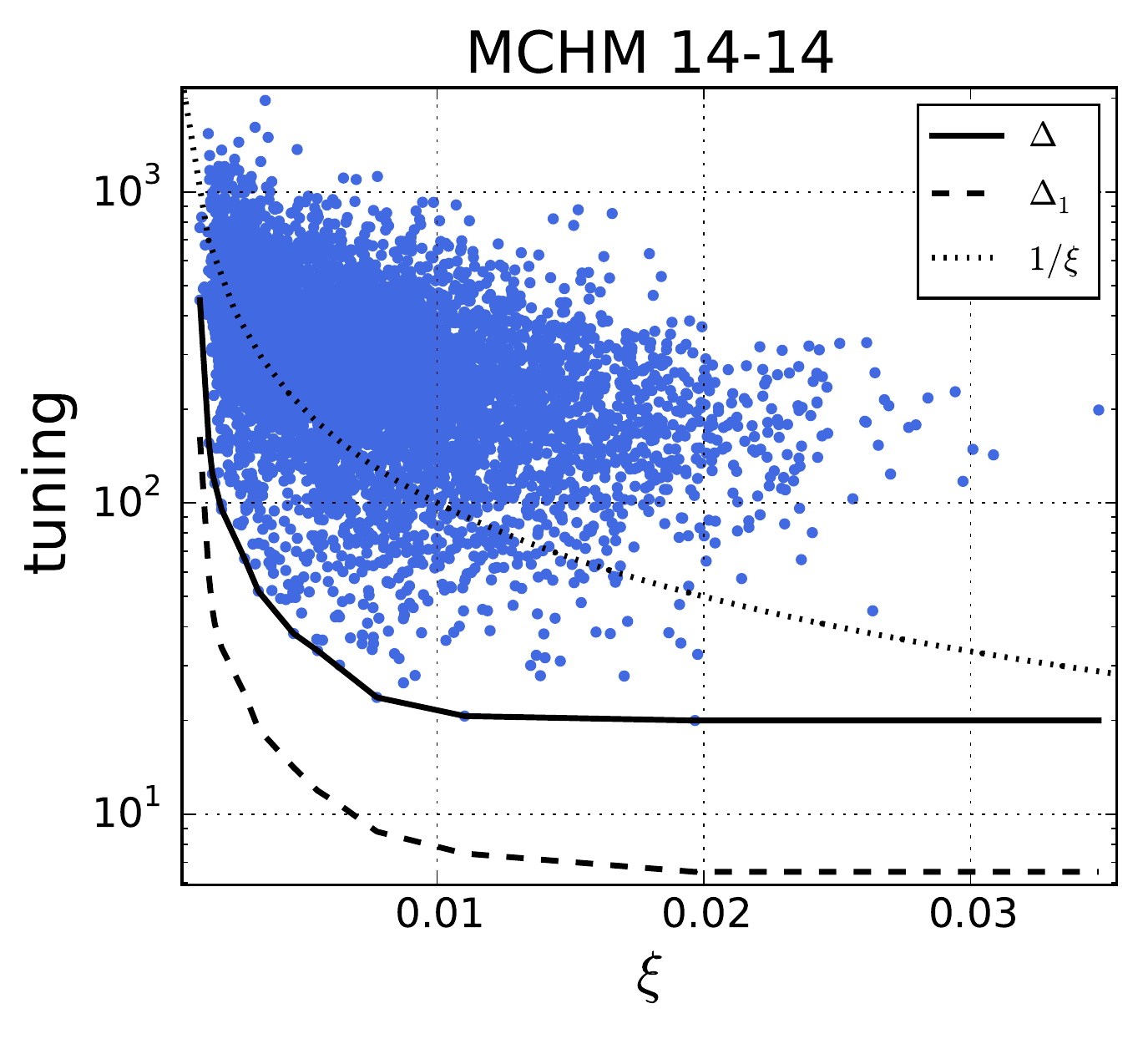}\vspace{-5mm}
\end{center}
\caption{Tuning in the 14-14 model as a function of $\rho$ mass, top partner mass, Higgs couplings and $\xi$. Further details are given in the text.\label{fig:1414r}}
\end{figure}

Full results for the 14-14 model are shown in figure~\ref{fig:1414r}. The tuning is a little less severe than in the 5-5 model, particularly for higher top partner masses, and the minimum we find is $\Delta=20$, i.e.\ 5\%. The 14-14 models tend to have a much higher compositeness scale and, therefore, much more SM-like Higgs couplings (note that $r_b\neq r_t$ in this model and only the minimum tuning for $r_b$ is shown). Lighter top partners in the ${\bf2}_{7/6}$ SM multiplet are still preferred suggesting that the composite sector is not too strongly coupled: $m_T\sim g_Tf$ so the composite sector coupling, $g_T$, cannot be too large. A very high degree of elementary-composite mixing in the top quark sector is preferred, at the limit of the range of viability in our effective theory approach. Interestingly the naive, $1/\xi$ estimate of tuning actually overestimates the tuning here.

The tuning can be most constrained in this model by improving limits on the top partner mass. Even then the change in the constraint is mild; a limit $\Delta\gtrsim30$ is reached for top partners heavier than 2 TeV\@. In the Higgs coupling sector a precision better than 2\% is required to outperform this. Hence the 14-14 model will remain relatively natural for the foreseeable future.

\subsection{14-1 model results}

\begin{figure}
\begin{center}
\includegraphics[width=0.47\textwidth]{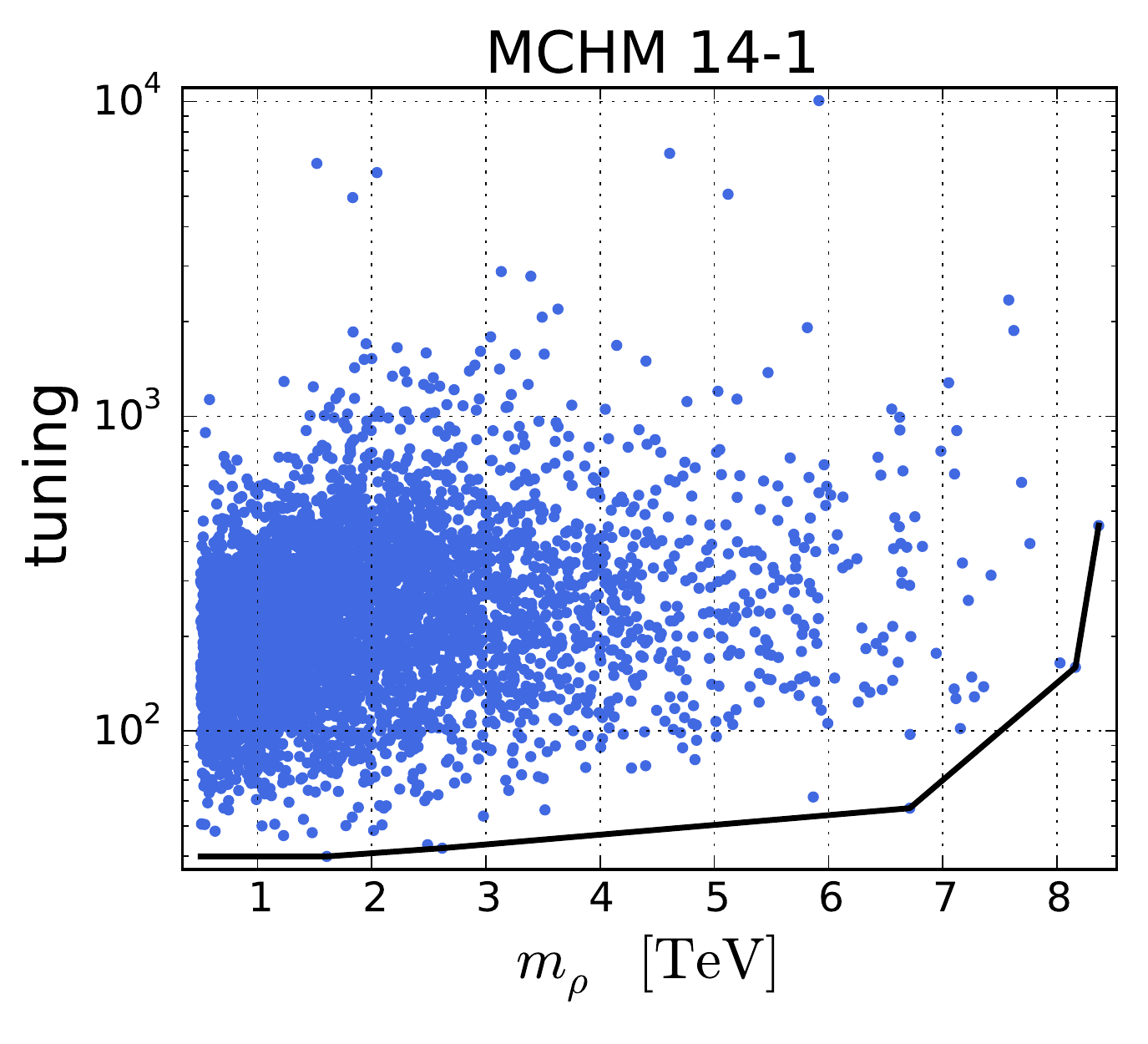}\hspace{5mm}
\includegraphics[width=0.47\textwidth]{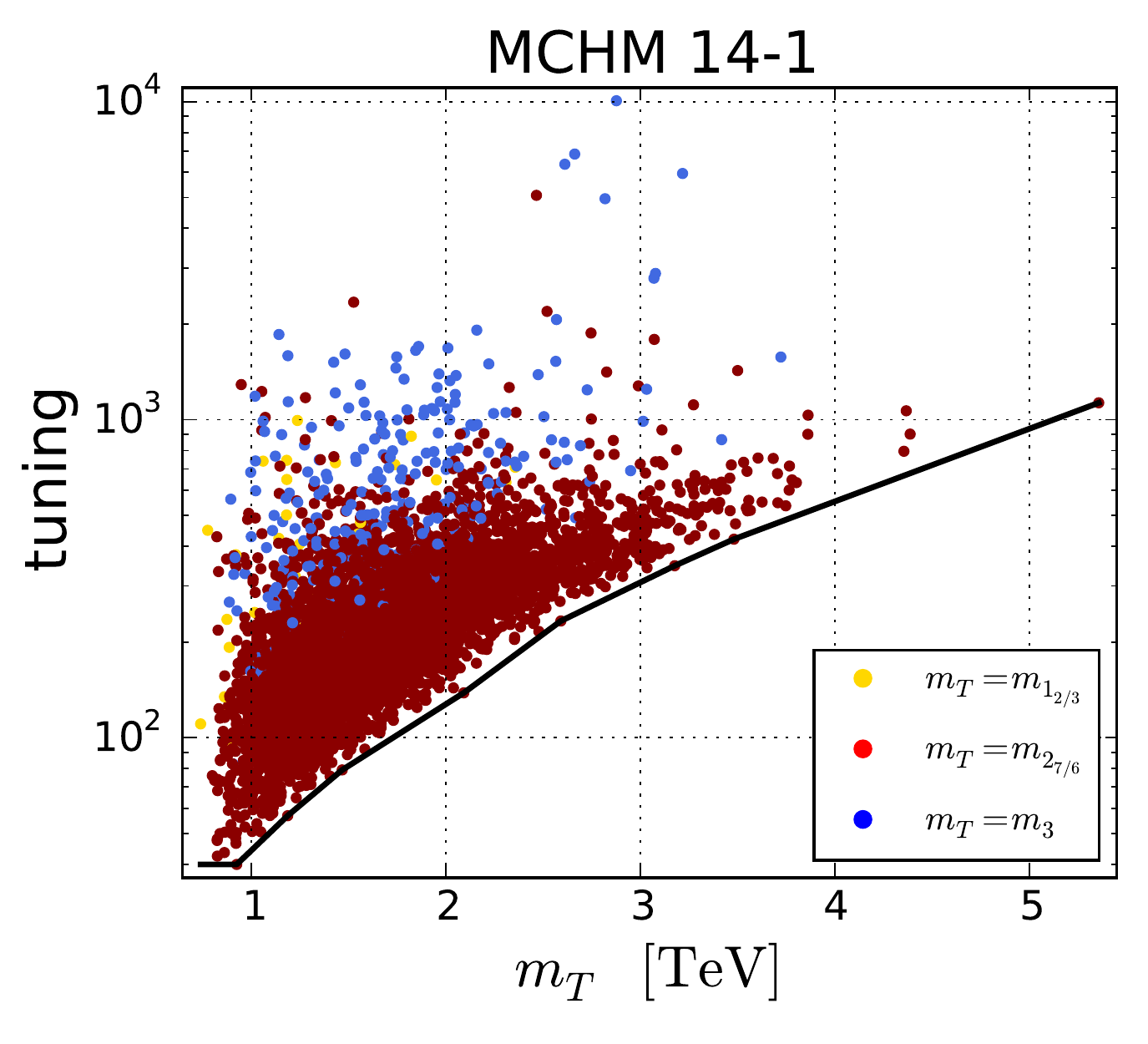}\\
\includegraphics[width=0.47\textwidth]{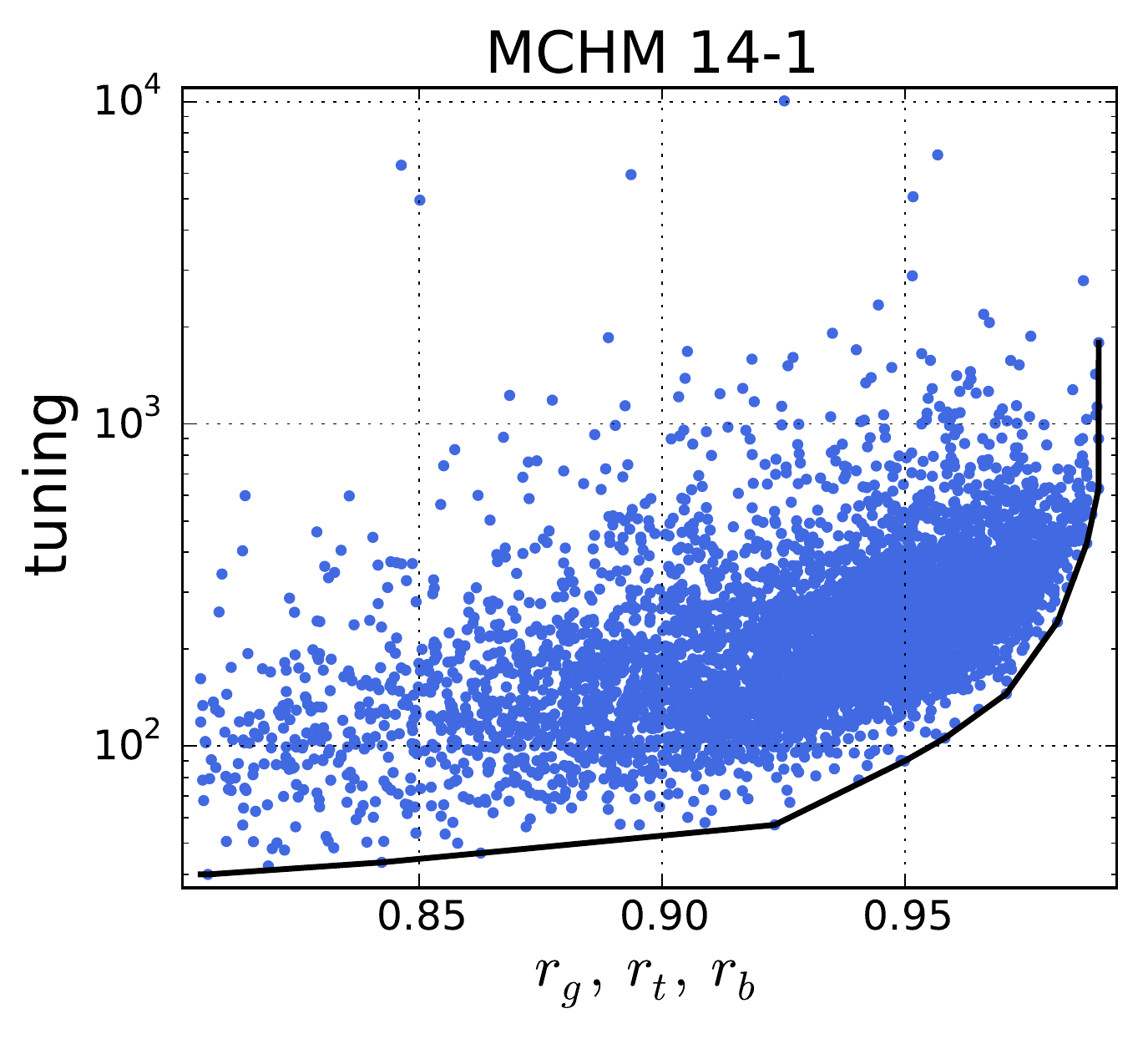}\hspace{5mm}
\includegraphics[width=0.47\textwidth]{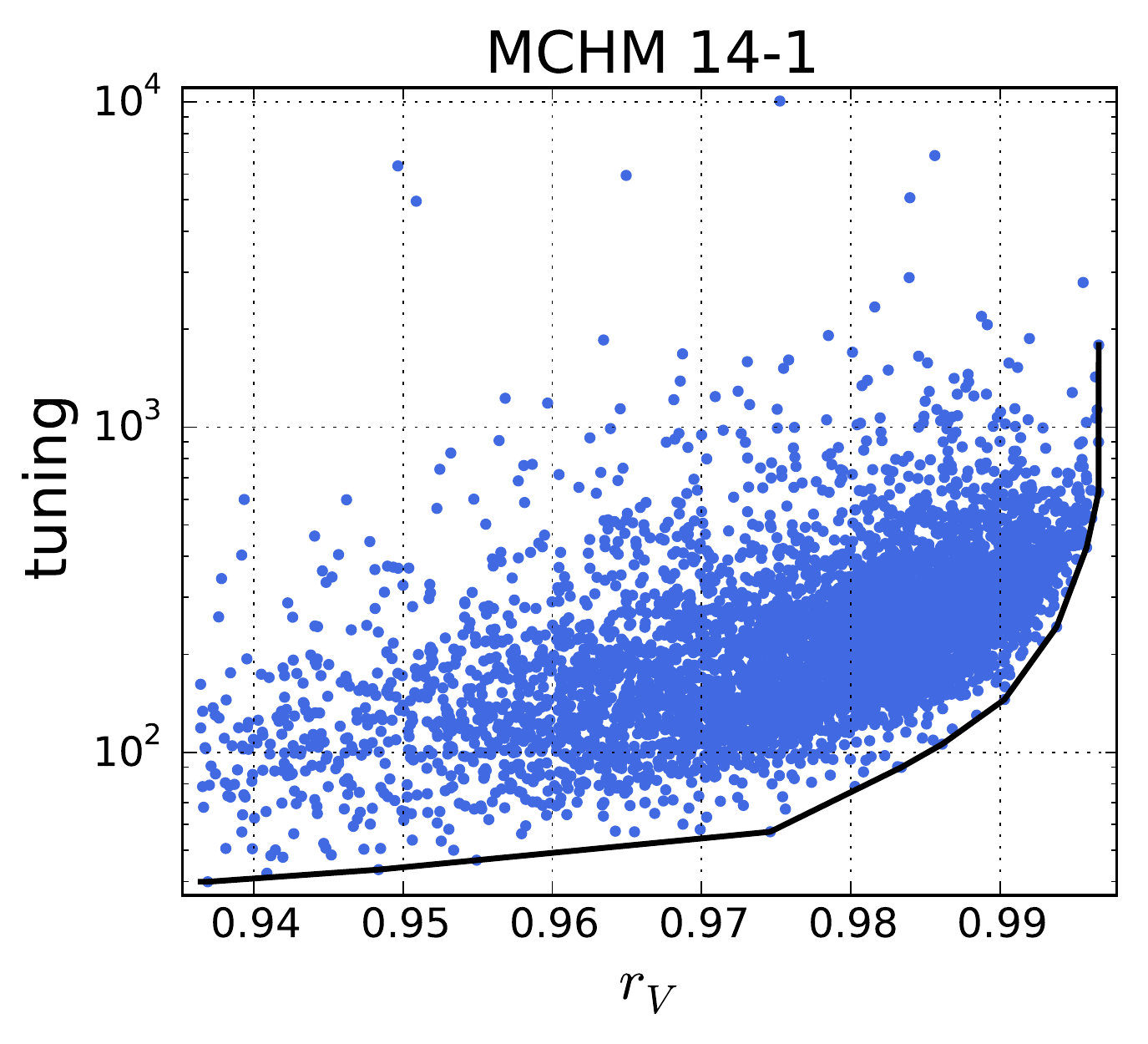}\\
\includegraphics[width=0.47\textwidth]{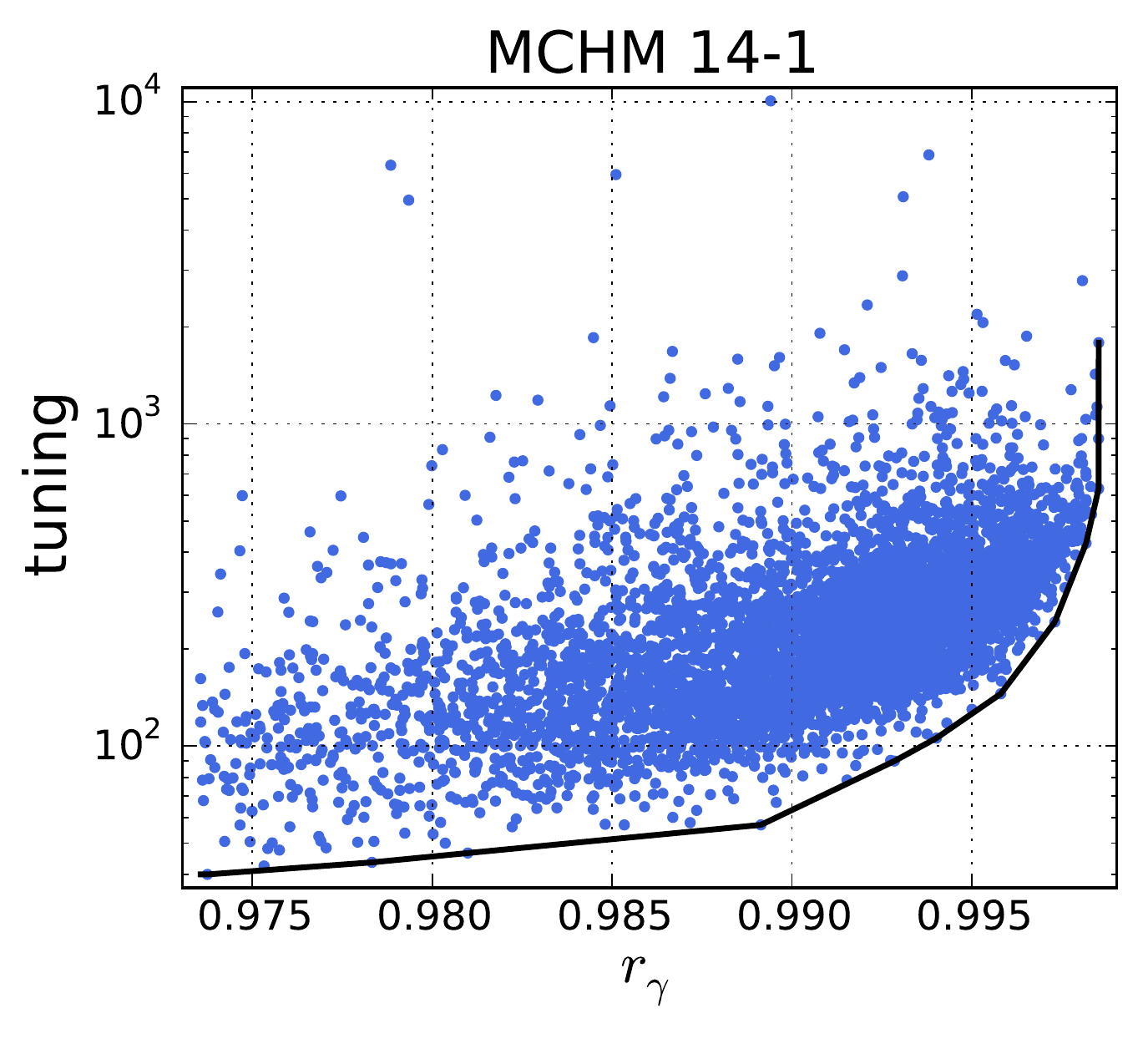}\hspace{5mm}
\includegraphics[width=0.47\textwidth]{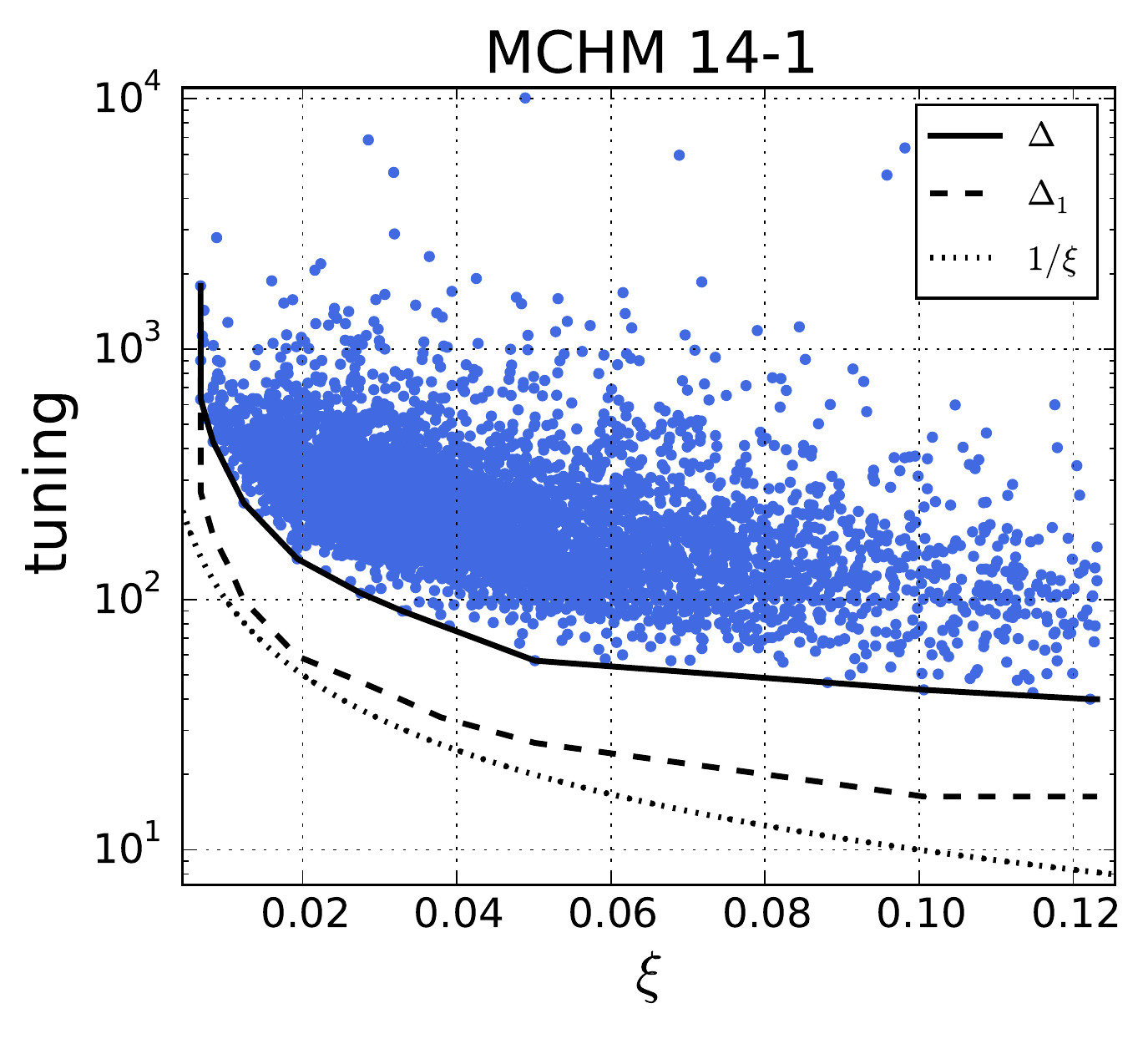}\vspace{-5mm}
\end{center}
\caption{Tuning in the 14-1 model as a function of $\rho$ mass, top partner mass, Higgs couplings and $\xi$. Further details are given in the text.\label{fig:141r}}
\end{figure}

Full results for the 14-1 model are shown in figure~\ref{fig:141r}. The tuning behaves similarly to that of the 5-5 model, albeit starting from a slightly higher base. The minimum we find is $\Delta=40$, i.e.\ 2.5\%. The compositeness scale tends to be lower than in the 14-14 model so constraints from Higgs coupling measurements are more important. Even so the best way to constrain the tuning in this model is via top partner searches, specifically for those in the ${\bf2}_{7/6}$ SM multiplet, and excluding top partners lighter than 2 TeV pushes the tuning to $\Delta\gtrsim130$. A precision better than 3\% in Higgs coupling measurements is required to outperform this.

\subsection{Discussion}

A common feature of all models studied here is that the severity of the tuning is being driven by constraints on the top partner masses, for the immediate future at least. After 300~fb$^{-1}$ of 14~TeV collisions at the LHC top partners lighter than 2~TeV are expected to be excluded~\cite{1311.0299,ColliderReach}. Comparing this with the expected precision to which the Higgs couplings can be measured with the same data set, around 9\% for $hWW$ and worse for the other couplings~\cite{1403.7191}, and it is clear that searching for top partners is a much more powerful way to probe naturalness.

In the more distant future this may change. A higher energy proton-proton collider could constrain top partner masses even further; lower limits of around 5~TeV and 9.5~TeV can be expected after 3000~fb$^{-1}$ of 33~TeV and 100~TeV collisions respectively~\cite{1311.0299,ColliderReach}, leading to tunings worse than 0.1\% in the models studied here. However, similar constraints on the tuning can be expected from a electron-positron collider sat on the Higgs resonance. TLEP~\cite{1308.6176}, for example, would measure the $hZZ$ coupling to an accuracy of about~0.15\%, implying a tuning worse than 0.1\% in the 5-5 and 14-1 models (due to its high compositeness scale top partner searches remain more powerful in the 14-14 model).

The main advantage of using an electron-positron collider to probe naturalness in this context is that it would be able to place meaningful constraints on a wider variety of models. Constraints coming from top partner searches rely on a connection between the Higgs mass and the masses of the top partners. This holds for the models studied here, where the large top quark Yukawa is explained by top quark compositeness, but other classes of model do exist. A composite right-handed tau can change this conclusion~\cite{1410.8555} as can the use of colourless states to stabilise the composite Higgs mass, like in composite twin Higgs models~\cite{0811.0394}, for example. A proper quantification of the tuning in such models is beyond the scope of this work, but we expect that the relationship between modifications to the Higgs couplings and the tuning will be qualitatively similar to the models that we have studied.

\section{Conclusions\label{sec:conclusions}}

Double tuning -- i.e.\ independent tunings to simultaneously get the right Higgs VEV and mass -- is important in composite Higgs models and should be accounted for. This can be done by using a tuning measure like the one we construct in section~\ref{sec:tuning}. Despite this extra source of tuning minimal 4D models based on an $SO(5)\to SO(4)$ symmetry breaking pattern remain fairly natural, the current minimum tunings being 3.7\%, 5\% and 2.5\% in the 5-5, 14-14 and 14-1 models. If top partners lighter than 2~TeV are excluded the tuning worsens to around 1\%, 3.3\% and 0.8\% respectively. We arrived at these values after performing a comprehensive scan of the full parameter space using a nested sampling algorithm, then applying the latest collider constraints on top partner masses and Higgs coupling deviations. Other, more model dependent constraints can be applied so our values provide a conservative minimum.

\section*{Acknowledgements}

We thank Peter Athron, Noel Dawe and Tony Gherghetta for helpful comments and discussions. This work was supported by the Australian Research Council. MJW is supported by the Australian Research Council Future Fellowship FT140100244. JB is grateful to the CERN theory division for its hospitality and partial support during the completion of this work.

\newpage
\appendix

\section{Model details\label{app:DMD}}

The models we consider consist of two sites -- an elementary site and a composite site. All SM degrees of freedom other than the Higgs live on the elementary site~\cite{DeCurtis:2011yx,1106.2719}. The composite site comprises a non-linear sigma model describing the pNGB Higgs and a set of vector-boson and fermion states representing the resonances. The two sites are connected via a set of link fields described by a second non-linear sigma model. The link fields are responsible for mixing between the two sites and for giving masses to the vector bosons on the composite site.

Initially the elementary site has a global symmetry $G_{\rm e}=SO(5)\times U(1)_X$ and the composite site has a gauge symmetry $G_{\rm c}=SO(5)\times U(1)_X$. The gauge fields associated with $G_{\rm c}$ are our vector-boson states, $\rho$ and $\rho_X$. The overall $G_{\rm e}\times G_{\rm c}$ symmetry is then spontaneously broken to the global, diagonal $G=SO(5)\times U(1)_X$ subgroup. The NGBs associated with this symmetry breaking are our link fields, $\Omega$. Since $G_{\rm c}$ is gauged the link fields are eaten by the $G_{\rm c}$ gauge fields to produce massive, vector-boson resonances transforming in the adjoint representation of $G$. At the same time $G_{\rm c}$ is spontaneously broken to an $H_{\rm c}=SO(4)\times U(1)_X$ subgroup on the composite site. The NGBs associated with the $G_{\rm c}\to H_{\rm c}$ symmetry breaking, $\Sigma$, will go on to provide our Higgs.

Upon including the SM degrees of freedom on the elementary site $G$ is explicitly broken to the usual, $SU(2)_L\times U(1)_Y$ electroweak symmetry of the standard model. This is precipitated by gauging only the $SU(2)_L\times U(1)_Y$ subgroup of $G_{\rm e}$ and by embedding the SM fermions in incomplete representations of $G_{\rm e}$. Nonetheless, the explicit breaking is weak (couplings and masses on the composite site are much greater than their elementary site counterparts) so the symmetry structure discussed above remains approximately intact and the Higgs, embedded in $\Sigma$, remains light due to its pNGB nature.

The Higgs is embedded into the spurion
\begin{align}\label{eq:SigmaDef}
\Sigma=e^{i\pi/f}\begin{pmatrix} 0 & 0 & 0 & 0 & 1 \end{pmatrix}=\begin{pmatrix} 0 & 0 & 0 & s_h & c_h \end{pmatrix}
\end{align}
where $\pi\equiv\pi^aX^a$ contains the four NGBs for the four broken $SO(5)$ generators, $\{X^a\}$, $s_h\equiv\sin(h/f)$ and $c_h\equiv\cos(h/f)$.  The NGB decay constant is denoted by $f$ and we work in the unitary gauge such that $[\pi^a]=(0,0,0,h)$.  The spurion transforms as a ${\bf5}$ of $SO(5)$ and acts a convenient book-keeping device when determining the interactions of the pNGB Higgs allowed by the symmetries of the model.

The Lagrangian for these models splits into contributions from the elementary site, the composite site and the mixing terms. The contribution from the elementary site is
\be
\cL_{\rm e}=-\frac{1}{4g_{2,{\rm e}}^2}W_{\mu\nu}W^{\mu\nu}-\frac{1}{4g_{1,{\rm e}}^2}B_{\mu\nu}B^{\mu\nu}+\pfrac{\Lambda}{d_qm_Q}^2\bar{q}i\slashed{D}_{\rm e}q+\pfrac{\Lambda}{d_tm_T}^2\bar{t}^ci\slashed{D}_{\rm e}t^c+\ldots
\ee
where $W$ and $B$ are the usual $SU(2)_L$ and $U(1)_Y$ field strength tensors; $q$ is a (two-component) fermion for the third-generation, left-handed quark doublet; $t^c$ is a fermion for the right-handed top quark, and $D_{\rm e}$ is the covariant derivative involving the elementary gauge fields. Dots denote terms involving the lighter fermions and, for later convenience, the quark kinetic terms are not canonically normalised -- the normalisation factor will be explained shortly.

The contribution from the composite site is
\begin{align}
\cL_{\rm c}={} & {}-\frac{1}{4g_\rho^2}\rho_{\mu\nu}\rho^{\mu\nu}-\frac{1}{4g_X^2}\rho_{X,\mu\nu}\rho_X^{\mu\nu}+\frac{f_{\rm c}^2}{2}(D_{{\rm c},\mu}\Sigma)(D_{\rm c}^\mu\Sigma)^T+\bar{Q}i\slashed{D}_{\rm c}Q+\bar{Q}^ci\slashed{D}_{\rm c}Q^c \\
& {}+\bar{T}i\slashed{D}_{\rm c}T+\bar{T}^ci\slashed{D}_{\rm c}T^c-m_QQQ^c-m_TTT^c-m_YQT^c-Y(\Sigma)QT^c+\mbox{h.c.}+\ldots \nonumber
\end{align}
where $\rho$ and $\rho_X$ are the field strength tensors for the composite, $G_{\rm c}$ gauge field; $\Sigma$ is the real scalar \eqref{eq:SigmaDef} containing the Higgs fields, and $D_{\rm c}$ is the covariant derivative involving the composite gauge fields. $Q$ and $T$ are (two-component) fermions, each coming in a Dirac pair. These fermions have diagonal mass terms, $m_Q$ and $m_T$, an off-diagonal mass term, $m_Y$, (assuming it is consistent with $G_{\rm c}$, which it is in the models studied here) and Yukawa-like terms, $Y(\Sigma)$, coupling the fermions to the Higgs via the spurion. The forms of the Yukawa-like terms depend on which representations of $G_{\rm c}$ the fermions belong to. Note that $Q^cT$ terms are not present despite being allowed by all symmetries of the model. These terms are omitted to ensure that the Higgs potential remains finite.

Finally, the mixing terms are
\be
\cL_{\rm m}=\frac{f_\Omega^2}{4}(D_{{\rm e+c},\mu}\Omega)(D_{\rm e+c}^\mu\Omega)^\dag+\Lambda\sbrack{R_q(\Omega)qQ^c+R_t(\Omega)t^cT}+\mbox{h.c.}+\ldots
\ee
where $\Omega$ is the complex scalar parameterising the $G_{\rm e}\times G_{\rm c}\to G$ NGBs\@. These transform under both $G_{\rm e}$ and $G_{\rm c}$ so the covariant derivative, $D_{\rm e+c}$, contains both elementary and composite gauge fields. The subsequent terms then use $\Omega$ to mix $q$ and $t^c$ with $Q^c$ and $T$ in a way respecting the initial $G_{\rm e}\times G_{\rm c}$ symmetry (or at least they would if $q$ and $t^c$ came in complete $G_{\rm e}$ representations). This is done by using projections, $R(\Omega)$, corresponding to the representations of $G_{\rm e}$ that $q$ and $t^c$ are embedded in. Owing to the non-canonical normalisation of the elementary fermions the actual couplings associated with the mixing terms for $q$ and $t^c$ go like $d_qm_Q$ and $d_tm_T$ respectively. The common scale, $\Lambda$, is cancelled in most expressions for physical quantities, although persists as an independent variable in the 14-1 model in which $d_t$ is also absorbed into $\Lambda$

In the end we are mostly interested in the low-energy effective theory coming from the above Lagrangian, which can be derived by integrating out degrees of freedom from the composite site: $\rho$, $Q$ and $T$. Recall that the link fields are eaten by the $\rho$'s so will not appear in the effective theory either. The detailed structure on the composite site gets encoded in momentum-dependent (i.e.\ non-local) terms in the effective theory's Lagrangian. In momentum space this Lagrangian is
\begin{align}\label{eq:Leff}
\cL_{\rm eff}={} & {}\frac{1}{2}P^T_{\mu\nu}\sbrack{\Pi_W(p^2,h)W_\mu W_\nu+\Pi_B(p^2,h)B_\mu B_\nu+\Pi_{WB}(p^2,h)W^3_\mu B_\nu}+{} \nonumber\\
& {}\Pi_q(p^2,h)\bar{q}\slashed{p}q+\Pi_t(p^2,h)\bar{t}^c\slashed{p}t^c+M(p^2,h)tt^c+\mbox{h.c.}
\end{align}
where $\Pi$ and $M$ denote the form factors encoding the effects of the composite sector and $P^T$ is the transverse projection operator.  Explicit expressions for the form factors can be derived once the embeddings for the elementary fermions have been chosen.

The one-loop Higgs potential is found to be
\begin{align}\label{eq:Vint}
V(h)={} & {}\int_0^\infty\frac{{\rm d}p^2}{16\pi^2}p^2\nbrack{\frac{9}{2}\ln\sbrack{\Pi_W(p^2,h)}} \nonumber\\
& {}-2N_c\int_0^\infty\frac{{\rm d}p^2}{16\pi^2}p^2\ln\sbrack{p^2(1+\Pi_q(p^2,h))(1+\Pi_t(p^2,h))+|M(p^2,h)|^2}
\end{align}
keeping only the contributions from the $SU(2)_L$ gauge fields and the top quark. It is more usually expanded in powers of $s_h$ to give
\be\label{eq:Vexp}
V(h)=-\gamma s_h^2+\beta s_h^4.
\ee
The Higgs VEV and mass found from this potential are
\begin{align}
\xi & =\frac{\gamma}{2\beta} & m_h^2 & =\frac{8\beta}{f^2}\xi(1-\xi)
\end{align}
and \eqref{eq:Leff} yields a top quark mass
\be
m_t=\frac{M(0,v)}{\sqrt{\Pi_q(0,v)\Pi_t(0,v)}}.
\ee

\subsection{Gauge sector variables and form factors}

Form factors in the gauge sector depend only on the symmetry breaking pattern so are the same in all models studied here. We vary $g_\rho$, $f_{\rm c}$ and $f_\Omega$ via the mixing angle and masses
\begin{align}
t_\theta & =\frac{g_{2,{\rm e}}}{g_\rho} &
m_\rho^2 & =\frac{1}{2}g_\rho^2f_{\rm c}^2 &
m_a^2 & =\frac{1}{2}g_\rho^2(f_{\rm c}^2+f_\Omega^2).
\end{align}
The form factor for the $W$ boson is
\be
\Pi_W=-\frac{p^2(p^2-(1+t_\theta^2)m_\rho^2)}{g_2^2(1+t_\theta^2)(p^2-m_\rho^2)}+\frac{1}{4}s_h^2\sbrack{\frac{2m_\rho^2(m_a^2-m_\rho^2)t_\theta^2}{g_2^2(1+t_\theta^2)(p^2-m_a^2)(p^2-m_\rho^2)}}
\ee
where $g_2$ is the observed $SU(2)_L$ gauge coupling. Plugging into \eqref{eq:Vint} and performing the integral results in a contribution to the $s_h^2$ coefficient in \eqref{eq:Vexp}
\be
\gamma_g=-\frac{9m_\rho^4(m_a^2-m_\rho^2)t_\theta^2}{64\pi^2(m_a^2-(1+t_\theta^2)m_\rho^2)}\ln\sbrack{\frac{m_a^2}{(1+t_\theta^2)m_\rho^2}}
\ee
at leading order in $t_\theta$. There is no equivalent contribution to the $s_h^4$ coefficient and higher orders terms in $s_h$ or $t_\theta$ will not be considered.

\subsection{5-5 model variables, form factors and masses\label{app:55}}

This model was developed in detail in ref.~\cite{DeCurtis:2011yx}. The form factors can conveniently be expressed in terms of the functions
\begin{align}
\widehat{\Pi}_5(m_1,m_2,m_3) & =\frac{(m_2^2+m_3^2-p^2)\Lambda^2}{p^4-(m_1^2+m_2^2+m_3^2)p^2+m_1^2m_2^2} \nonumber\\
\widehat{M}_5(m_1,m_2,m_3) & =\frac{m_1m_2m_3\Lambda^2}{p^4-(m_1^2+m_2^2+m_3^2)p^2+m_1^2m_2^2}
\end{align}
as
\begin{align}
\Pi_q & =\frac{\Lambda^2}{d_q^2}+\widehat{\Pi}_5(m_Q,m_T,m_Y)+\frac{1}{2}s_h^2\sbrack{\widehat{\Pi}_5(m_Q,m_T,m_Y+Y)-\widehat{\Pi}_5(m_Q,m_T,m_Y)} \nonumber\\
\Pi_t & =\frac{\Lambda^2}{d_t^2}+\widehat{\Pi}_5(m_T,m_Q,m_Y+Y)+s_h^2\sbrack{\widehat{\Pi}_5(m_T,m_Q,m_Y)-\widehat{\Pi}_5(m_T,m_Q,m_Y+Y)} \nonumber\\
M & =\frac{1}{\sqrt{2}}s_hc_h\sbrack{\widehat{M}_5(m_Q,m_T,m_Y+Y)-\widehat{M}_5(m_Q,m_T,m_Y)}
\end{align}
Plugging into \eqref{eq:Vint} and performing the integral results in contributions to the $s_h^2$ and $s_h^4$ coefficients in \eqref{eq:Vexp} at, respectively, leading (quadratic) and subleading (quartic) order in $d_t$ and $d_q$. Higher orders terms in $s_h$ or $d_q$ and $d_t$ will not be considered. The top partner masses, up to small, electroweak corrections, are found as follows
\begin{itemize}
\item $p^2=m_{{\bf1}_{2/3}}^2$ is a zero of $\Pi_t$ at $s_h=0$
\item $p^2=m_{{\bf2}_{1/6}}^2$ is a zero of $\Pi_q$ at $s_h=0$
\item $p^2=m_{{\bf2}_{7/6}}^2$ is a pole of $\widehat{\Pi}_5(m_Q,m_T,m_Y)$.
\end{itemize}

\subsection{14-14 model variables, form factors and masses\label{app:1414}}

This model was developed in detail in ref.~\cite{Panico:2012uw} (see also ref.~\cite{Carena:2014ria}). The form factors can conveniently be expressed in terms of the functions
\begin{align}
\widehat{\Pi}_{14}(m_1,m_2,m_3,m_4) & =\frac{(m_1^2+m_2^2-p^2)\Lambda^2}{p^4-(m_2^2+m_3^2+m_4^2)p^2+m_3^2m_4^2} \nonumber\\
\widehat{M}_{14}(m_1,m_2,m_3) & =\frac{m_1m_2m_3\Lambda^2}{p^4-(m_1^2+m_2^2+m_3^2)p^2+m_1^2m_2^2}
\end{align}
as
\begin{align}
\Pi_q={} & {}\frac{\Lambda^2}{d_q^2}+\widehat{\Pi}_{14}(m_T,m_Y+\tfrac{Y_1}{2},m_Q,m_T)+{} \nonumber\\
& {}\frac{5}{4}s_h^2\left[\widehat{\Pi}_{14}(m_T,m_Y,m_Q,m_T)-2\widehat{\Pi}_{14}(m_T,m_Y+\tfrac{Y_1}{2},m_Q,m_T)\right. \nonumber\\
& \phantom{\frac{5}{4}s_h^2\left[\widehat{\Pi}_{14}(m_T,m_Y,m_Q,m_T)\right.}\left.{}+\widehat{\Pi}_{14}(m_T,m_Y+\tfrac{4Y_2}{5},m_Q,m_T)\right]-{} \nonumber\\
& {}\frac{1}{4}s_h^4\left[3\widehat{\Pi}_{14}(m_T,m_Y,m_Q,m_T)-8\widehat{\Pi}_{14}(m_T,m_Y+\tfrac{Y_1}{2},m_Q,m_T)\right. \nonumber\\
& \phantom{4s_h^4\left[3\widehat{\Pi}_{14}(m_T,m_Y,m_Q,m_T)\right.}\left.{}+5\widehat{\Pi}_{14}(m_T,m_Y+\tfrac{4Y_2}{5},m_Q,m_T)\right]
\end{align}
\begin{align}
\Pi_t={} & {}\frac{\Lambda^2}{d_t^2}+\widehat{\Pi}_{14}(m_Q,m_Y+\tfrac{4Y_2}{5},m_Q,m_T)+{} \nonumber\\
& {}\frac{5}{2}s_h^2\sbrack{\widehat{\Pi}_{14}(m_Q,m_Y+\tfrac{Y_1}{2},m_Q,m_T)-\widehat{\Pi}_{14}(m_Q,m_Y+\tfrac{4Y_2}{5},m_Q,m_T}+{} \nonumber\\
& {}\frac{5}{16}s_h^4\left[3\widehat{\Pi}_{14}(m_Q,m_Y,m_Q,m_T)-8\widehat{\Pi}_{14}(m_Q,m_Y+\tfrac{Y_1}{2},m_Q,m_T)\right. \nonumber\\
& \phantom{16s_h^4\left[3\widehat{\Pi}_{14}(m_Q,m_Y,m_Q,m_T)\right.}\left.{}+5\widehat{\Pi}_{14}(m_Q,m_Y+\tfrac{4Y_2}{5},m_Q,m_T)\right]
\end{align}
\begin{align}
M={} & {}\frac{\sqrt{5}}{2}s_hc_h\sbrack{\widehat{M}_{14}(m_Q,m_T,m_Y+\tfrac{4Y_2}{5})-\widehat{M}_{14}(m_Q,m_T,m_Y+\tfrac{Y_1}{2})}-{} \nonumber\\
& {}\frac{\sqrt{5}}{8}s_h^3c_h\left[3\widehat{M}_{14}(m_Q,m_T,m_Y)-8\widehat{M}_{14}(m_Q,m_T,m_Y+\tfrac{Y_1}{2})\right. \nonumber\\
& \phantom{\sqrt{5}s_h^3c_h\left[3\widehat{M}_{14}(m_Q,m_T,m_Y)\right.}\left.{}+5\widehat{M}_{14}(m_Q,m_T,m_Y+\tfrac{4Y_2}{5})\right].
\end{align}
Plugging into \eqref{eq:Vint} and performing the integral results in contributions to both the $s_h^2$ and $s_h^4$ coefficients in \eqref{eq:Vexp} at leading (quadratic) and subleading (quartic) order in $d_t$ and $d_q$. Higher orders terms in $s_h$ or $d_q$ and $d_t$ will not be considered. The top partner masses, up to small, electroweak corrections, are found as follows
\begin{itemize}
\item $p^2=m_{{\bf1}_{2/3}}^2$ is a zero of $\Pi_t$ at $s_h=0$
\item $p^2=m_{{\bf2}_{1/6}}^2$ is a zero of $\Pi_q$ at $s_h=0$
\item $p^2=m_{{\bf2}_{7/6}}^2$ is a pole of $\widehat{\Pi}_{14}(m_T,m_Y+\tfrac{Y_1}{2},m_Q,m_T)$
\item $p^2=m_{\bf3}^2$ is a pole of $\widehat{\Pi}_{14}(m_T,m_Y,m_Q,m_T)$.
\end{itemize}

\subsection{14-1 model variables, form factors and masses\label{app:141}}

This model was developed in detail in ref.~\cite{Panico:2012uw} (see also ref.~\cite{Carena:2014ria}). There are two ${\bf14}$'s, $Q_1$ and $Q_2$, with separate mass terms and no $T$ as the fully-composite $t^c$ is already able to play this role.\footnote{Without the second ${\bf14}$ an accidental symmetry causes a part of the form factor crucial for reducing the tuning in this model to vanish.} Hence $d_t$ is absorbed into $\Lambda$ and $\Lambda$ is treated as the independent variable. The form factors can conveniently be expressed in terms of the functions
\begin{align}
\widehat{\Pi}_5(m_1,m_2,m_3) & =\frac{(m_2^2+m_3^2-p^2)\Lambda^2}{p^4-(m_1^2+m_2^2+m_3^2)p^2+m_1^2m_2^2} \nonumber\\
\widehat{\Pi}_1(m_1,m_2,m_3) & =\frac{(m_2^2-p^2)\Lambda^2}{p^4-(m_1^2+m_2^2+m_3^2)p^2+m_1^2m_2^2} \nonumber\\
\widehat{M}_1(m_1,m_2,m_3) & =\frac{m_1(m_2^2-p^2)\Lambda^2}{p^4-(m_1^2+m_2^2+m_3^2)p^2+m_1^2m_2^2}
\end{align}
as
\begin{align}
\Pi_q={} & {}\frac{\Lambda^2}{d_q^2}+\widehat{\Pi}_5(m_{Q_1},m_{Q_2},m_Y+\tfrac{Y_1}{2})+{} \nonumber\\
& {}\frac{5}{4}s_h^2\left[\widehat{\Pi}_5(m_{Q_1},m_{Q_2},m_Y)-2\widehat{\Pi}_5(m_{Q_1},m_{Q_2},m_Y+\tfrac{Y_1}{2})\right. \nonumber\\
& \phantom{\frac{5}{4}s_h^2\left[\widehat{\Pi}_5(m_{Q_1},m_{Q_2},m_Y)\right.}\left.{}+\widehat{\Pi}_5(m_{Q_1},m_{Q_2},m_Y+\tfrac{4(Y_1+Y_2)}{5})\right]-{} \nonumber\\
& {}\frac{1}{4}s_h^4\left[3\widehat{\Pi}_5(m_{Q_1},m_{Q_2},m_Y)-8\widehat{\Pi}_5(m_{Q_1},m_{Q_2},m_Y+\tfrac{Y_1}{2})\right. \nonumber\\
& \phantom{\frac{1}{4}s_h^4\left[3\widehat{\Pi}_5(m_{Q_1},m_{Q_2},m_Y)\right.}\left.{}+5\widehat{\Pi}_5(m_{Q_1},m_{Q_2},m_Y+\tfrac{4(Y_1+Y_2)}{5})\right]
\end{align}
and
\begin{align}
\Pi_t & =1+\widehat{\Pi}_1(m_{Q_1},m_{Q_2},\Lambda) \nonumber\\
M & =\widehat{M}_1(m_{Q_1},m_{Q_2},\Lambda).
\end{align}
Plugging into \eqref{eq:Vint} and performing the integral results in contributions to both the $s_h^2$ and $s_h^4$ coefficients in \eqref{eq:Vexp} at leading (quadratic) and subleading (quartic) order in $d_q$. Higher orders terms in $s_h$ or $d_q$ will not be considered. The top partner masses, up to small, electroweak corrections, are found as follows
\begin{itemize}
\item $p^2=m_{{\bf1}_{2/3}}^2$ is a zero of $\Pi_t$ at $s_h=0$
\item $p^2=m_{{\bf2}_{1/6}}^2$ is a zero of $\Pi_q$ at $s_h=0$
\item $p^2=m_{{\bf2}_{7/6}}^2$ is a pole of $\widehat{\Pi}_5(m_{Q_1},m_{Q_2},m_Y+\tfrac{Y_1}{2})$
\item $p^2=m_{\bf3}^2$ is a pole of $\widehat{\Pi}_5(m_{Q_1},m_{Q_2},m_Y)$.
\end{itemize}

\bibliographystyle{JHEP-2}
\bibliography{MCHMTuning}
\end{document}